\documentclass{article}

\usepackage[noadjust]{cite}
\usepackage[english]{babel} 
\usepackage[utf8]{inputenc} 
\usepackage[a4paper,left=3cm,bottom=3cm]{geometry} 
\usepackage{indentfirst} 
\usepackage{enumerate, enumitem} 
\usepackage{amsmath} 
\usepackage{bm} 
\usepackage{graphicx} 
\usepackage{multicol} 
\usepackage[export]{adjustbox} 
\usepackage{pdflscape} 
\usepackage{fancyhdr} 
\usepackage{hyperref}
\hypersetup{colorlinks,allcolors=blue}
\usepackage{flafter} 
\usepackage[framemethod=tikz]{mdframed} 
\usepackage{color} 
\usepackage{wrapfig}
\usepackage{float} 
\usepackage{lipsum}
\usepackage{xcolor, xurl}

\graphicspath{ {Images/} }  

\title{White Paper on Critical and Massive Machine Type Communication Towards 6G}
\author{\textbf{Editors:} 
Nurul H. Mahmood\thanks{N. H. Mahmood (\textit{Corresponding editor, email: nurulhuda.mahmood@oulu.fi}), O. Lopez, K. Mikhaylov, C.-F. Liu, E. Annanper{\"a}, H. Alves and P. Sepp{\"a}nen are with University of Oulu, Finland. 
I. Moerman is with imec - Ghent University, Belgium. 
O.-S. Park, E. Kim, J. Shin, G.-Y. Park, S.-K. Kim and C. Yoon are with ETRI, South Korea. 
E. Mercier and J.-B. Dore are with CEA-Leti, France. 
A. Munari, F. Clazzer, H. Bartz and G. Liva are with German Aerospace Center (DLR), Germany. 
S. B{\"o}cker and C. Wietfeld are with TU Dortmund, Germany. 
R. J{\"a}ntti and M. Andraud are with Aalto University, Finland. 
R. Pragada is with InterDigital, USA. 
Y. Ma is with ZTE, China. 
Y. Chen is with Huawei Technologies, Canada. 
E. Garro is with Universitat Polit\`{e}cnica de Val\`{e}ncia, Spain. 
Y. Sadi is with Kadir Has University, Turkey. 
F. Burkhardt is with Fraunhofer IIS, Germany. 
K. Anwar is with Telkom University, Indonesia.}, 
Onel Lopez, 
Ok-Sun Park, \\
Ingrid Moerman, Konstantin Mikhaylov,
Eric Mercier, 
Andrea Munari, \\
Federico Clazzer, 
Stefan B{\"o}cker and 
Hannes Bartz. \\
\textbf{Contributors:} Nurul H. Mahmood, Stefan B{\"o}cker, Andrea Munari, \\
Federico Clazzer, Ingrid Moerman, Konstantin Mikhaylov, Onel Lopez, \\
Ok-Sun Park, Eric Mercier, Hannes Bartz, Riku J{\"a}ntti, \\
Ravikumar Pragada, Yihua Ma, Elina Annanper{\"a}, Christian Wietfeld, \\
Martin Andraud, Gianluigi Liva, Yan Chen, Eduardo Garro, \\
Frank Burkhardt, Hirley Alves, Chen-Feng Liu, Yalcin Sadi, \\
Jean-Baptiste Dore, Eunah Kim, JaeSheung Shin, Gi-Yoon Park, \\
Seok-Ki Kim, Chanho Yoon, Khoirul Anwar and Pertti Sepp{\"a}nen}
\date{\today}

\begin{document}

\pagenumbering{gobble} 
\maketitle
\tableofcontents
\clearpage

\pagenumbering{arabic} 
\setcounter{page}{1}


\renewcommand{\abstractname}{Executive Summary}
\begin{abstract}

The society as a whole - including vertical sectors such as industries, intelligent transport systems, smart cities, etc. - is becoming increasingly digitalized. Machine Type Communication (MTC), encompassing its massive and critical aspects, and ubiquitous wireless connectivity are among the main enablers of such digitization at large. 

The recently introduced 5G New Radio is natively designed to support both aspects of MTC to promote the digital transformation of the society and particularly improve the overall efficiency of different vertical sectors. However, it is evident that some of the more demanding requirements cannot be fully supported by 5G networks. Alongside, further development of the society towards 2030 will give rise to new and more stringent requirements on wireless connectivity in general, and MTC in particular. 

Driven by the societal trends towards 2030, the next generation (6G) will be an agile and efficient convergent network serving a set of diverse service classes and a wide range of key performance indicators (KPI). 

This white paper explores the main drivers and requirements of an MTC-optimized 6G network, and presents a set of research directions for different aspects of MTC that can be synthesized through the following six key research questions: 
\begin{itemize}
    \item Will the main KPIs of 5G, namely reliability-latency-scalability, continue to be the dominant KPIs in 6G; or will emerging metrics like energy-efficiency, end-to-end (E2E) performance measures and sensing become more important?
    \item How to deliver different E2E service mandates with different KPI requirements through a multi-disciplinary approach jointly considering optimization at the physical up to the application layer?
    \item What are the key enablers towards designing ultra-low power receivers and highly efficient sleep modes to support ultra-low-cost ultra-low-power or even passive MTC devices?
    \item How to tackle a disruptive rather than incremental joint design of a massively scalable waveform and medium access policy to efficiently support global connectivity for MTC?
    \item How to support new service classes characterizing mission-critical and dependable MTC in 6G through multifaceted connectivity and non-cellular centric wireless solutions?
    \item What are the potential enablers of long term secure schemes considering the heterogeneous requirements and capabilities of MTC devices? How to design lightweight and flexible usable ways of handling privacy and trust in MTC by combining the user perspective with the technical perspective?
\end{itemize}
\end{abstract}

\newpage

\section{Introduction}
\label{sec:intro}

\subsection*{Background and Objective}
The year 2020 is expected to witness large scale global deployment of fifth generation (5G) wireless network. While 5G is being deployed across the globe, the research community has already started posing the question - \textit{What will the sixth generation (6G) be?} In fact, this was the main theme of the first 6G Wireless Summit held in Levi, Finland in March 2019, which led to the world's first 6G white paper published in September 2019~\cite{6gWP2019}. 

To delve further into this question, 6G Flagship project\footnote{\url{www.6gflagship.com}} plans to develop a set of 12 new White Papers exploring selected thematic topics in depth in conjunction with $2^{nd}$ 6G Wireless Summit held virtually in March 2020\footnote{\url{http://www.6gsummit.com/}}. The objective is to investigate key research questions and potential enabling technologies for 6G with respect to the different thematic topics. 

This white paper addresses the evolution of critical and massive machine type communication (MTC) towards 6G. In particular, the white paper will explore the societal development and use cases pertinent to MTC towards 2030; identify key performance indicators (KPI) and requirements; and discuss a number of potential technical enablers for MTC in future 6G networks. The main take-away points are highlighted through the six key research questions presented in the executive summary. 

\subsection*{State of The Art}

Wireless connectivity is now an ubiquitous utility, just like water and electricity. Wireless networks were initially designed to connect people. However, they have now evolved to enable machine type communication (MTC) allowing applications residing at different machine type devices (MTD) to interconnect wirelessly without the need for human intervention. 

A network formed by different connected MTDs is known as Internet of Things (IoT). IoT enables a wide range of applications with diverse requirements in many different verticals. This ranges from best-effort connectivity for simple sensors to high data-rate, highly reliable real-time connectivity, e.g., for vehicles. International Telecommunication Union (ITU) has classified IoT use cases into two broad service classes, namely ultra reliable low latency communication (URLLC) and massive MTC (mMTC)~\cite{ITU_2410}. 

5G New Radio (NR) network targets to optimize the resource utilization for high connection density (e.g., $1$ million connections per square kilometer) mMTC use cases, and also support a timely, efficient, and reliable mechanism to deliver the target information to one or multiple devices for URLLC use cases with $99.999\%$ reliability at $1$ millisecond (ms) user plane latency~\cite{ITU_2410}. 

The basic NR framework defined in 3GPP Release 15 provides scalable and configurable air interface design to support ultra-low latency transmission through configurable numerology and frame structures, and very short mini-slot with down to two symbols duration. Similarly, grant-free (GF) transmission has been specified in NR to reduce signaling overhead and latency, which is suitable for both URLLC and mMTC, especially in the uplink.

Future releases of 5G NR (Release 16 in mid-2020 and Release 17 later) will include further enhancements for URLLC and mMTC service. In addition to connectivity through the cellular infrastructure, sidelink connectivity with another IoT device or smartphone will be introduced. Moreover, GF transmission will be extended to support sidelink transmissions and be enhanced with non-orthogonal multiple access (NOMA). Finally, Narrow band IoT (NB-IoT) and LTE-M will be integrated with 5G NR to provide dedicated MTC services. Please refer to~\cite{GMB19:5Gevolution} and references therein for a detailed discussion. 

Though cellular networks offer access to diverse services and rates through larger bandwidths and a more extensive network, the supported devices and the network itself are usually too power-hungry and costly for many MTC applications. To alleviate these limitations, low power wide area networks (LPWAN) such as SigFox, LoRA/LoRAWAN, Ingenu, have been specifically designed for MTC. These are usually extremely power efficient and are able to provide connectivity over a long range (up to tens of kilometers), though the data rates and the supported use cases are rather limited. Alongside, there exists a number of proprietary industrial communication networks (ICN) designed to serve the need of factory and process automation in industries. 

While 5G NR and other wireless systems have enabled mMTC and URLLC services under certain scenarios, the true vision of IoT connectivity as required by the diverse range of MTC applications is yet to be realized. In particular, MTC is fundamentally different from conventional human type communication (HTC). The current approach of modifying the existing wireless system that are primarily designed for HTC to meet IoT connectivity needs has proven to be rather inefficient and unscalable. 

There is also a lack in considering end-to-end (E2E) aspects as part of the design, that is, full protocol stack from the physical layer (PHY) to the application layer (APP), and full connection path from APP to APP. Moreover, different MTC scenarios and applications may have very divergent requirements. Hence, connectivity solutions have to be application-aware and may need multiple radio access technologies (multi-RAT) to ensure robustness. 

The number of IoT connected devices is expected to grow three-fold in the next decade (from about $11$ billion\footnote{\url{https://www.ericsson.com/en/mobility-report/reports/november-2019/iot-connections-outlook}} in 2019 to $30$ billion by 2030\footnote{\url{https://www.statista.com/statistics/802690/worldwide-connected-devices-by-access-technology/}}), serving a wide variety of use cases with very diverse requirements. As 5G NR and other MTC systems continue to evolve in the near future, there is a need to design a robust, scalable, and efficient sixth generation (6G) wireless network that can address the limitations of the existing systems and meet the emerging requirements of 2030 society and beyond.  

Although 6G research is still in the exploration phase, there are already a good number of publications exploring what 6G will be. An initial vision of what 6G might be is presented in~\cite{KML18_6G}, which are further elaborated in~\cite{david_6g_2018, SBC19_6G, ZXM_19_6G, ZFW_19_6G, SBG_19_6G, viswanath6G_2020}. A comprehensive vision of 6G as a human-centric mobile communications network, along with issues beyond communication technologies that could hamper research and deployment of 6G are discussed in~\cite{DAS_19_whatShould6G}. Considering a more specific perspective, potential key enablers for MTC in 6G are discussed in~\cite{MAL_20_6GMTC}. 

This white paper intends to further contribute to the ongoing discussions shaping 6G design by presenting an unified and holistic picture of an intelligent and E2E-optimized wireless network capable of supporting MTC evolution towards the 6G era.

\section{MTC Megatrends Towards 2030}
\label{sec:requirements}

\begin{figure}[htb]
    \centering
    \includegraphics[width=0.99\columnwidth]{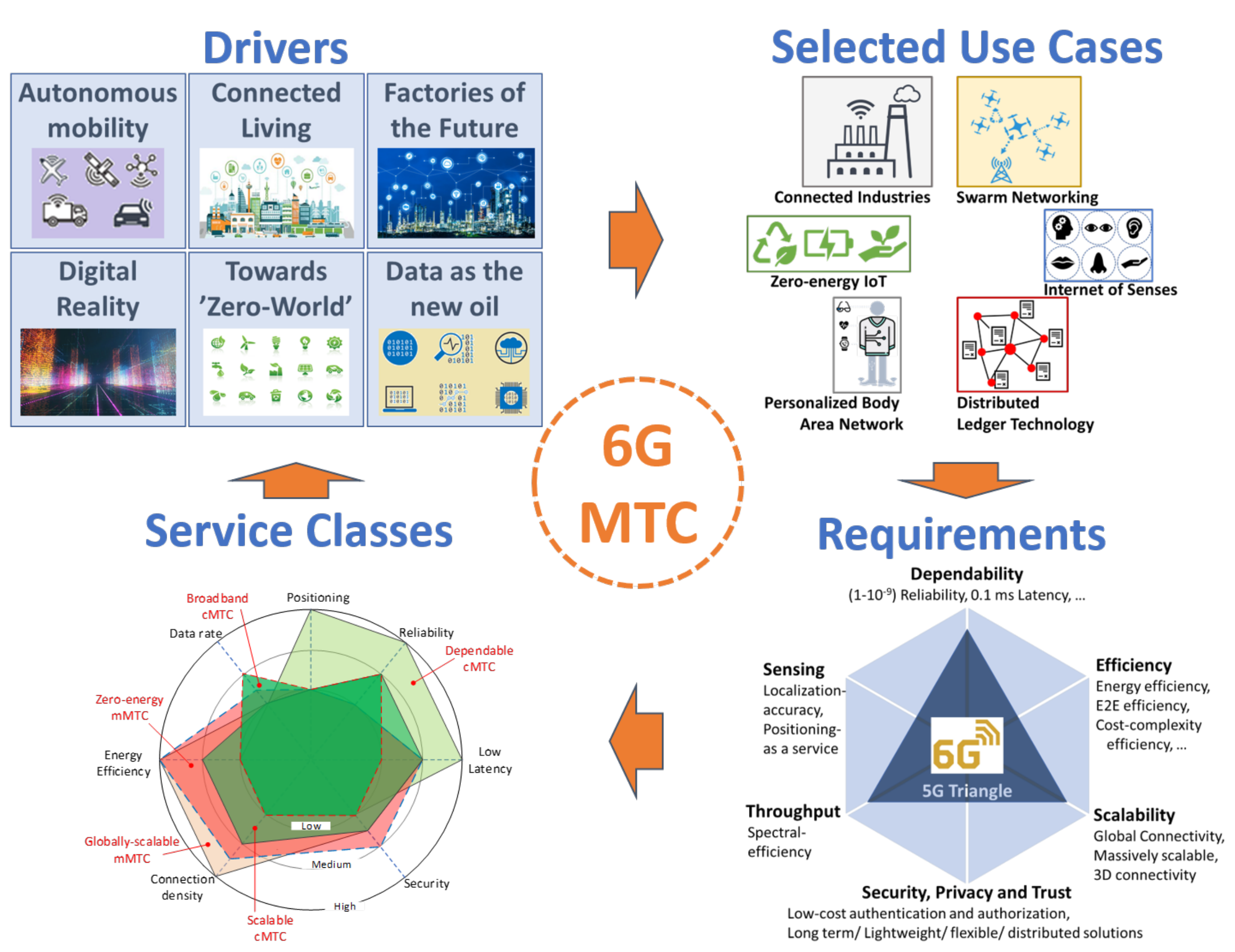}
    \caption{Drivers, use cases, requirements and service classes of 6G MTC towards 2030.}
    \label{fig:drivers}
\end{figure}

Our society is developing fast with many transformative changes taking place within a short span of time, for which MTC is a key technology enabler. The main drivers, some representative use cases, key requirements and service classes of MTC towards 2030 are schematically presented in Fig.~\ref{fig:drivers} and further detailed next. 

\subsection{Drivers}

Frost $\&$ Sullivan's Visionary Innovation Group\footnote{\url{https://ww2.frost.com/research/visionary-innovation/}} has identified 12 Global Megatrends futurecasting key themes that will shape the society at large by 2030. Out of these megatrends, those that we believe will strongly drive the evolution of MTC in the next decade are listed below.

\subsubsection*{Autonomous mobility}
MTC will be a key driver in the growing trend towards making anything that moves autonomous and intelligent. For instance, Intelligent Transport Systems (ITS) demand provision of reliable connectivity between multiple actors. Improving the level of driving automation (from level 3 to level 4 or level 5) is expected to increase the amount of sensors and edge processing systems integrated within the vehicles. To be most efficient, these systems will utilize real-time wireless connectivity to its sensor/actuator nodes and external processing units. 
The increasing role of artificial intelligence (AI) in both the vehicle and the infrastructure also imposes more strict requirements on the information transfer process. Additionally, the recent advances in Unmanned Vehicles and related applications will require a three-dimensional (3D) connectivity landscape. 
    
\subsubsection*{Connected Living}
Ubiquitous connectivity anytime and anywhere will facilitate new possibilities at home, in entertainment, health and work, and in providing citizen services. 
6G will drive our cities to be super smart and fully connected with a plethora of autonomous services, which
 will be enabled by a forecasted $30$ billion connected IoT devices by 2030. 
    
\subsubsection*{Factories of the Future}
Transformational shifts in production and industries and novel human-machine interactions are enabled by MTC. By supporting diverse needs and requirements, it will provide the groundwork towards Industry 5.0. Industry 5.0 targets producing customized and personalized products in mixed sensing/actuation/haptics scenarios instead of the dominant data collection and analytics scenarios of Industry 4.0. Industry 5.0 will involve much more interactivity, both local and remote, with even more diverse and stringent demands regarding wireless connectivity compared to Industry 4.0~\cite{Nahavandi_I5dot0_2019}. 

\subsubsection*{Digital reality as frontier technology}    
Augmented/ virtual/ mixed reality (XR) will evolve from a niche  market application to widespread adoption resulting in new use cases and applications. MTC will play a fundamental role in facilitating the user experience and enabling novel and efficient man-machine interfaces to present the data coming from machines in a more natural way. Rapid advances in Wireless Brain-Computer Interactions and Multisensory XR requirements will drive a massive proliferation of cost-effective miniaturized smart wearable and implant sensors with Quality of Service (QoS) guarantees.

\subsubsection*{Towards `Zero' World}
The proliferating vision of a zero world concept is shifting focus towards technologies that innovate to zero, such as for example `zero-energy' technologies and `zero-touch' systems. Particularly, zero-energy requirements will require MTC to deliver higher performance at zero or very low energy consumption; or to provide zero-latency and zero-error capabilities for enabling real-time control and Emergency IoT use cases.
    
\subsubsection*{Data as the new oil}
Towards 6G, data markets will emerge connecting data suppliers and customers. The data generated by the widely distributed MTDs will have enormous business and societal values. The value-added services of data marketplaces will be empowered by emerging technologies like AI and Blockchain, while adding new data-centric KPIs like age of information (AoI), privacy and localization accuracy. 

All the above societal and technological trends are fundamental for realizing United Nations' 2030 Sustainable Development Goals (SDG)\footnote{\url{www.un.org/development/desa/disabilities/envision2030.html}} for an inclusive, trustworthy and self-sustainable society. 


\subsection{Use Cases}

Considering the foreseen economic, societal, technological, and environmental context of the 2030 era, future network demands must be jointly met in a holistic fashion. It is therefore not straightforward to forecast use cases of MTC towards 6G. However, in the following, we list a number of plausible use cases, which by their generality and complementarity, we believe depict a representative picture of the different possible MTC use cases in 6G. 

\subsubsection*{Connected Industries} 
Industry 4.0 converges advanced manufacturing techniques with data-driven technologies and AI tools to improve the operational and performance efficiency of enterprises. Further evolution towards Industry 5.0 and the emerging trend of personalization and customization will require future factories to be even more agile and adaptable, supported by massive connectivity of mobile and versatile production assets. This requires the evolution of existing KPIs of reliability, latency and throughput, while also supporting new ones like localization accuracy and those measuring man-machine interface performance.

\subsubsection*{Swarm Networking}
Autonomous vehicles like self-driving cars, automated guided vehicles (AGV) and unmanned aerial vehicles (UAV) have started finding application in a myriad of use cases. A swarm is a group of such devices collectively performing tasks in a distributed fashion to achieve an overall mission objective. Towards 2030 such swarms will be commonplace in shop floors, connected logistics and transport, emergency response, etc, requiring robust connectivity solutions for such complex MTC networks.

\subsubsection*{Personalized Body Area Network}
Wearable devices like smart watches and ear-buds are a part-and-parcel of our everyday life today. By 6G era, such MTDs will radically evolve, thanks to advances in man-machine interface that will render the devices seamlessly integrated, for example in our clothing or even implanted as skin-patches and bio-implants, while being effortless to operate~\cite{viswanath6G_2020}. While 6G MTC networks will play a significant role as enabler of such personalized body area network, different aspects of it will also impose different requirements on the network. 

\subsubsection*{Zero-energy IoT}
Right at the edge or extreme-edge of the IoT ecosystem machines are mostly low-power sensors. Such MTDs are powered by batteries or energy harvesters and are very limited in computing and storage capabilities to reduce costs and enlarge lifetime. Circuit technology advances towards 2030 aim at reducing their power consumption up to the point of keeping them \textit{perpetually} alive \cite{Portilla.2019}. 
Wireless Energy Transfer (WET) and backscatter technology are fundamental enablers. Lifetime requirements would demand more than $40$ years of continuous operation, for which stand-by and active power consumption may require to be bellow $1$ nanowatt (nW) and $1$ microwatt ($\mu$W), respectively.

%

\subsubsection*{Internet of Senses}
Internet of Senses will allow all human senses to interact with machines by enabling haptic interaction with sensory or perceptive feedback. This will revolutionize the way we manipulate and interact with our surroundings, and will enable truly immersive steering and control in remote environments. Ultrareliable, ultra broadband and ultraresponsive near real-time network connectivity may be mandatory, demanding massive, specialized and adaptive sensor deployments and interactions.

\subsubsection*{Distributed Ledger Technology}
Distributed Ledger Technologies (DLT) allow value transactions between parties through decentralized trust. Towards 2030, MTC networks will expand DLT's application horizons because of the increasing need to transfer valuable, authenticated sensor data, services, or micropayments between the IoT devices and other parties. For instance, two cars meeting on the road may want to buy data  from each another about the road conditions ahead \cite{Elsts.2018}. These distributed sensing services demand connectivity with a synergistic mix of URLLC and mMTC to guarantee low-latency, reliable connectivity, and scalability.



\subsection{Requirements}
As evidenced by the above use cases, 6G KPIs and requirements will be diverse and include novel metrics alongside the existing ones considered in 5G. 

As a part of the 5G requirements for IMT-2020~\cite{ITU_2410}, ITU has emphasized a number of KPIs, such as reliability, latency, connection density, and energy efficiency (EE). In 6G, more stringent requirements will be inevitable for these KPIs, while the QoS demands will evolve to be E2E. For example, in industrial scenarios, closed-loop control applications will require E2E reliability of up to $1-10^{-7}$ to maintain close synchronization at E2E latencies as low as $1$ ms~\cite{5GACIA_connectedIdustries}, for which a per-link reliability of around $1-10^{-9}$ and user plane latency around $0.1$ms may be required. This would allow instant optimization based on real-time monitoring of sensors and the performance of components, collaboration between a new generation of robots, and the introduction of wirelessly connected wearables and augmented reality on the shop floor. 

Similarly, the emerging need for ultra  dense deployment of industrial IoT devices will require 3D connectivity supporting up to $100$ connections per $m^3$. In terms of EE, the ultimate 6G vision is zero-energy MTDs, achieved through a combination of efficient hardware design and energy harvesting techniques~\cite{david_6g_2018}.

Alongside the existing ones, new KPIs are becoming increasingly relevant for 6G. The main focus here is on AoI, interoperability, dependability, positioning, sustainability and E2E EE. AoI characterizes the freshness of information, and allows answering how frequently the information status at a sink node needs to be updated through status update transmissions from a source node. Hence, it is crucial for networked monitoring and control systems like cyber-physical systems. 

Considering the heterogeneity of access technologies (wired and wireless) and deployment scenarios, 6G technologies are expected to be capable of seamless integration and interoperability across the heterogeneous networks. Meanwhile, reliability will evolve to dependability, which is an umbrella QoS term characterizing the attributes of availability, reliability, security and system integrity used to characterize system life cycles and failures. Localization accuracy will be relevant for emerging applications using positioning as a service, such as controlling AGVs in factory floors. 

Finally, while 5G mostly addresses device EE, sustainability, cost (production, installation, maintenance and operational costs) and recently E2E EE (including energy consumption of the infrastructure), the EE vision towards 6G will be wider and more stringent. For instance, the total cost and energy consumption per successfully delivered bit at the application level between end devices including its environmental impact, will be of utmost importance towards the 6G era.

\subsection{MTC Service Classes}

MTC in 5G is split into URLLC, or critical MTC (cMTC), in controlled environments with small-payloads and low-data rates, and mMTC for large/dense deployments with sporadic traffic patterns. In the coming decade, this will evolve to even more stringent and heterogeneous constraints because of the emerging industrial use cases and the verticalization of the service provision, hence, demanding multi-dimensional optimization and scalable design. So 6G needs to serve very diverse applications ranging from data-rate hungry holographic images and connected 360 XR to massive access for various types of IoT devices. MTC service classes in 6G can thus be classified as follows: 

\begin{description}
    \item[Dependable cMTC,] which refers to supporting extreme ultra reliability and low latency along with other measures of dependability (e.g., security), as well as precise positioning, and can be seen as the direct evolution of 5G URLLC, e.g. for autonomous driving. 
    
    \item[Broadband cMTC,] which refers to supporting mobile broadband (MBB) data with high reliability and low latency, e.g. XR, cloud gaming, robotic aided surgery.
    
    \item[Scalable cMTC,] which refers to supporting massive connectivity with high reliability and low latency, e.g. critical medical monitoring and factory automation.

    \item[Globally-scalable mMTC,] which refers to supporting ultra-wide network coverage throughout all space dimensions, including volumetric density of devices. The role of UAV swarms and non-terrestrial networks (NTN) is fundamental towards global mMTC connectivity.
    
    \item[Zero-energy mMTC,] which covers massive deployments of EE radios with very long battery life and network lifetime, e.g. soil monitoring and precision agriculture. This aims to cover techniques for zero-energy radios, energy harvesting, backscatter communication and extreme EE resource allocation. 
\end{description}

\section{Potential Enabling Technologies}
\label{sec:enablers}

Novel approaches in the designing the enabling technology components for MTC towards 6G is necessary to successfully meet the 6G visions and requirements laid out in Section~\ref{sec:requirements}. In this section, we discuss a number of proposed solutions spanning across different layers of the protocol stack, ranging from efficient hardware design to the considerations at the applications layer itself. 

Holistic MTC network architecture presented in Section~\ref{sub:holistic} provides a bird's eye view of the solution landscape and frames the forthcoming discussion. Energy efficient hardware considerations like zero-energy air interface are presented next in Section~\ref{sub:sustainable}, followed by a discussion on enablers for globally available and massively scalable MTC services in Section~\ref{sub:scalable}.

Finally, mission-critical MTC serving the needs of industrial sector and related verticals, and privacy and security aspects considering the heterogeneity of MTC devices and applications are discussed in Sections~\ref{sub:cMTC} and~\ref{sub:privacy}, respectively.

\subsection{Holistic MTC Network Architecture}
\label{sub:holistic}

\subsubsection{Introduction to MTC Networks}
Owing to the sheer variety of scenarios, services and requirements, today's portfolio of the MTC RATs and the architectural options underlying the different MTC networks is excessively multitudinous and diverse. However, today there is no single killer MTC RAT – a ``Jack of all trades'' – which can address the needs of a decently significant share of envisaged MTC applications and use cases. It is also unlikely that such a RAT will appear in the near future. 

Likewise, not a single MTC RAT is omnipresent and, vice versa, in many environments, multiple RATs and networks do coexist, thus forming a heterogeneous multi-connectivity-enabled environment. However, such an environment is typically disjoint - the networks often belong to the different administrative domains – i.e., run by the various stakeholders, each of which manages its network independently of the others. Moreover, in the case of the RATs operating in the unlicensed bands (BLE, WiFi, IEEE 802.15.4 or LoRaWAN to give an example) – the different RATs and even multiple networks employing the same RAT may also compete for the same radio resources.  

The further development of the new MTC network architecture has to take off from this starting point to harmonize the connectivity landscape and make it sustainable, whilst maximizing the performance of the (i) individual machines, (ii) the network as a whole, and, in the end of the day, (iii) the plethora of services and applications served by the machines and the network(s).

\subsubsection{Key Trends}

A number of the current key trends affecting the MTC network architecture has to be accommodated in the new MTC network architecture towards 6G. These can be summarized as:

\begin{enumerate}
    \item New zero-energy air interfaces are emerging to support new class of devices that do not draw power from the MTD's battery but rather harvest the energy they need. The MTC network architecture needs to be evolved to accommodate these new devices that have drastically different requirements from the infrastructure compared to current devices. For instance, separate carriers could be needed to transfer power to the devices and to allow backscatter communications.  
    
    \item The evolution of networks from cell-based towards the cell-free. Cell-free networks imply no explicit connection established between a MTD and a base station (BS) of a network cell, thus reducing signaling overhead while improving the reliability by enabling multiple BSs to demodulate the incoming transmissions, either separately or jointly.
    
    \item Further increase of the network heterogeneity due to (i) the appearance of local micro-operators deploying new private or public networks (with ``crowd-driven networking'' paradigm representing one of the extremes), (ii) due to the introduction of the new RATs and the evolution of the existing RATs, and (iii) the shrinking of the cell sizes due to moving to higher frequency bands and/or applying more advance modulation and coding schemes leading to very geographically concentrated specialized deployments serving specific end goals.
    
    \item The enabling of the infrastructure-less and dynamically formed networks for critical missions, servicing out of cellular coverage and use cases implying high mobility (e.g., drone platooning or the mobile BS case).
    
    \item Finally, the need to support an ever growing number of very different traffic patterns and service classes (e.g., priorities, including emergency traffic) for individual and/or groups of MTDs.
\end{enumerate}

\subsubsection{Holistic MTC networks}
\begin{figure}[htb]
    \centering
    \includegraphics[width=0.9\columnwidth]{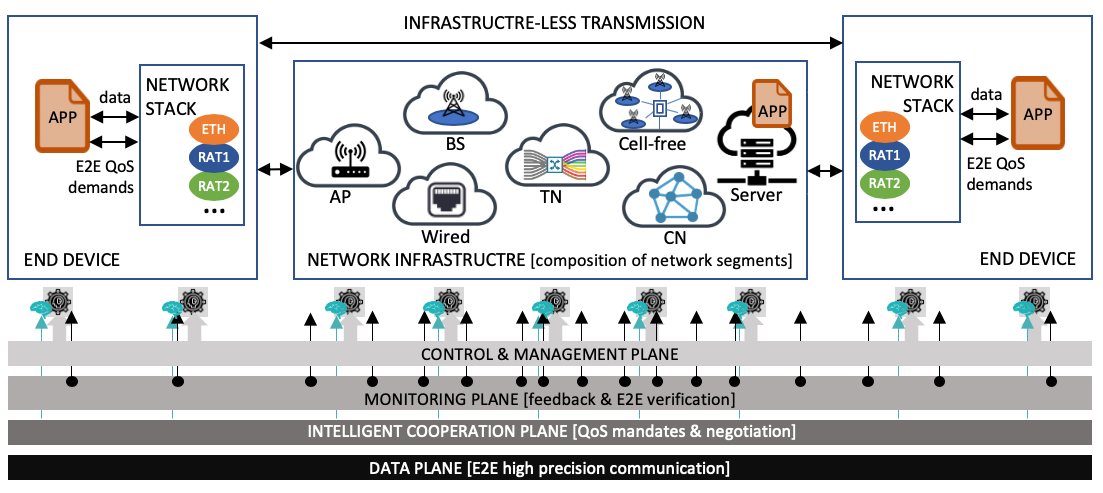}
    \caption{ Holistic MTC architecture driven by End-to-End QoS demands}
    \label{fig:Architecture}
\end{figure}

The envisaged holistic MTC network architecture (see Fig. ~\ref{fig:Architecture}) of the future should feature be highly efficient to fully utilize resources, intelligent at the edge for dynamic requirements and software-defined with iterable design. To increase the efficiency, the network requires the native support for dynamic and collaborative orchestration between communicating end-to-end applications across multiple heterogeneous network domains (wired/wireless/optical), operated by different stakeholders (cellular, public, private) involving different RATs. To facilitate interoperability, the interfaces, through which this should be handled, must be technology agnostic, meaning that end-to-end applications should not be aware on underlying technologies and RATs, but should still be able to express their requirements to the network. 

The selection of the RAT(s) and the possible configuration parameters need to be handled through open algorithms based on the end-to-end (E2E) QoS requirements of applications and the fairness considerations among them, which in some cases have to be made known to the network orchestrator(s), and mindful of the possible alternatives and the effects on the performance of the network as whole. The machine-learning methods represent a promising mechanism to help to address this challenge. 

Importantly, the set of the possible RAT alternatives, as well as the number and distribution of machine-clients, their traffic patterns and communication needs are not static and can change drastically both in space and in time. This postulates the need of equipping the network with efficient real-time localization and optimization mechanisms both of centralized and distributed nature, which can be deployed at the edge and even moved to the individual devices (e.g., as a mobile agent). 

The Introduction of the effectively manageable control and emergency traffic channels is also highly desirable. Importantly, the holistic MTC architecture should simultaneously be future proof and backward compatible not only within a single network operator domain but also across domains, supporting the interoperability between new and old generations and releases within one generation. In this context, softwarization of the network and radio functions, allowing for time- and cost-efficient update of the network and radio elements and new functionality introduction is especially crucial.

\subsubsection{Ultra-low power MTC networks}
Zero-energy devices require new air-interface which in turn has impact on the network architecture (cf. Section~\ref{sub:sustainable}). These devices require the network to illuminate them with relatively large power. The powering could be provided by using a dedicated carrier. Alternatively, information transmission of the base station or neighboring users could be utilized for this purpose. In the latter case, the backscatter devices form an underlay network within the MTC network consisting of active transmitters. Such an underlay operation is sometimes referred as ambient backscatter communication (AmBC), and is further discussed in Section~\ref{sub:scalable}. 

\subsubsection{Open Standards}
This is obvious that the design of a holistic post-5G MTC network architecture and development of the underlying optimization mechanisms and technical solutions introduces a major research challenge and requires the investment of substantial resources both by Academy and Industry. However, what we consider to be the most crucial challenge and the first step to be addressed on this way is reaching the agreement and developing the specification of the common interfaces and procedures allowing different networks and even their elements talking to each other. 

The next and even more arduous step, requiring consolidation of efforts and goodwill of the whole community as well as of the political and regulatory authorities, would be the development of these interface specifications into the universal cross-disciplinary standards spanning beyond the borders of individual regions and application domains. This would open the path to the development of the efficient, ubiquitous unplugged MTC in the context of 6G or, possibly, a later release.

\subsection{Energy Efficient MTC Devices}
\label{sub:sustainable}

In order to support a scalable deployment of massive MTDs in 6G, energy efficiency will be among the most important KPIs of mMTC. Efficient design of MTDs will be enabled through a combination of ultra-low power receiver architecture, enhanced energy harvesting techniques and zero energy communication schemes. On the other hand, collaborative and distributed intelligence at the device and the network are foreseen as a key enabler for cMTC. The first step in such an `intelligence-focused' design is the optimization of the device hardware itself through the development of `on device intelligence' blocks. This section discusses the design of energy efficient MTDs considering both, mMTC and cMTC use cases, though the discussion is more focused on the former.

\subsubsection{Ultra-Low Power Receiver Architecture}

Ultra low power (ULP) transceivers are essential to support the billions of expected MTDs in a sustainable way. While other aspects of MTD design developed significantly over the years, energy harvesting (EH) state of the art (SotA) shows that the available energy remains low and that no revolutionary technology appeared in the past decade. Hence sustainable mMTC operation cannot be solely powered through EH, including wireless power transfer; and further MTD development must still target ULP.

Significant power reduction can be achieved by considering the energy consumption of the MTD as a whole, instead of treating the individual elements, such as the antennas and RF transceivers, separately. The antennas should be part of the device package, and the interface with the RF transceiver must be considered, not only as a conventional impedance matching, but as an optimized intimate connection. Indeed, smaller radiating element, thereby with narrower bandwidth that has better selectivity and stronger robustness to blocking signals, can be beneficial with band filtering split along the whole chain, from the antenna to the embedded digital intelligence. 

The transceiver power consumption can also benefit from better power schemes of the various building blocks. Duty cycling or event-driven architectures are to be favored as they may directly impact the power consumption by the usage factor of the transceiver itself. It nevertheless necessitates the power-up settling time to be very short with high flexibility to enable swift change of state from sleep to on. 

Specifically considering the transmitter, the generated waveform and its amplification before emission has to comply with high linearity specification and energy efficiency, provided that the final power amplifier can handle flexible biasing mechanisms, like Envelop Tracking features, which are very efficient to make supply voltage fit the power amplifier requirements.

\begin{figure}[htb]
    \centering
    \includegraphics[width=0.99\columnwidth]{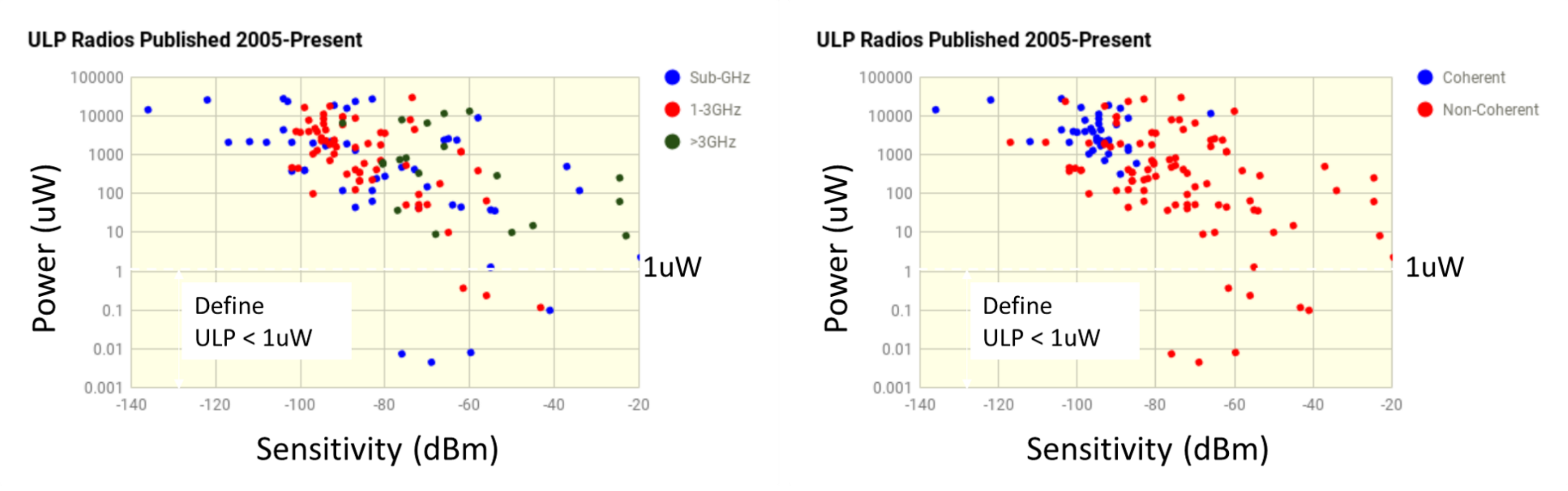}
    \caption{State of art ULP receiver design power consumption versus sensitivity per frequency band and type of design~\cite{wentzloffULPsurvey}.}
    \label{fig:ULP}
\end{figure}

Robustness in a wide-spectrum operation remains of high importance and the performance of the whole Rx chain has to optimized.  A graphical representation in Figure~\ref{fig:ULP} shows power consumption versus sensitivity~\cite{wentzloffULPsurvey}. It clearly demonstrates that sub-GHz band with simple non-coherent modulation schemes (on-off keying) represents a promising direction towards achieving below $100$ nW power consumption at less than $-80$ dBm sensitivity.


\subsubsection{Ambient Backscatter Communications for 6G mMTC}

\begin{figure}[htb]
    \centering
    \includegraphics[width=0.8\columnwidth]{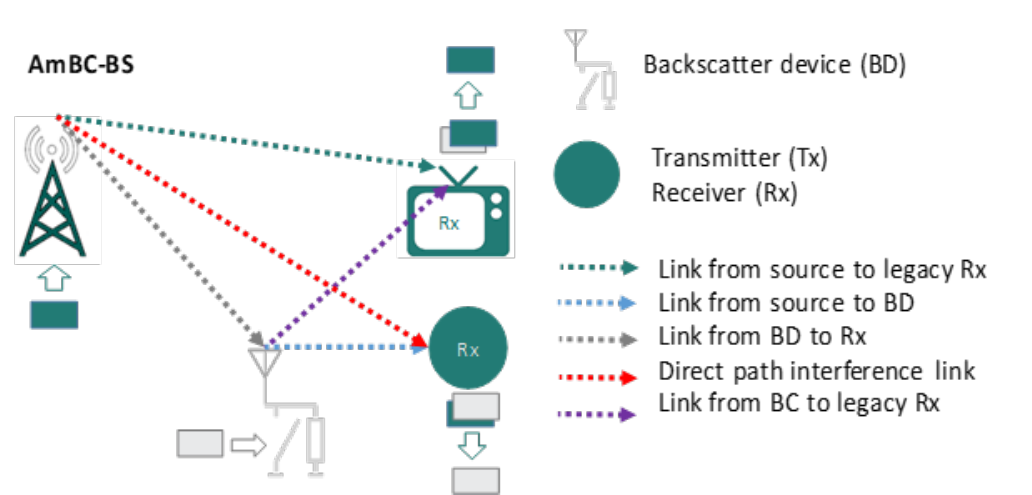}
    \caption{Deployment option for AmBC-BS.}
    \label{fig:ambc-bs}
\end{figure}

Ambient Backscatter Communications is a promising new technology to alleviate cost, power consumption and spectrum occupancy. In AmBC, backscatter devices can communicate with each other by modulating and reflecting received RF signals from ambient sources such as cellular BSs~\cite{Huynh2018}. Such method consumes significantly less power than typical active transmitter as no voltage-controlled oscillator and power amplifier are required. The key technical challenges in using AmBC for MTC are:

\begin{description}[align=left]
\item [Direct path interference -] The direct path signal power from the ambient source to the receiver is typically several order of magnitude stronger than the scattered path containing the information of interest. Interference cancellation (IC) can be applied but receivers with high dynamic range (high number of bits in the analog-to-digital converter) would be needed. Thus, a more efficient solution would be to cancel the signal in spatial domain using null steering. 

\item [Unknown and fast change of the amplitude and phase of the ambient signal -] AmBC receiver sees the amplitude and phase variations of the ambient signal as fast fading. For instance, legacy system using OFDM signal, looks like Rayleigh fading for the AmBC receiver. From the AmBC system perspective, the ideal ambient source signal would have constant amplitude. The case of multi-antenna receiver can use beamforming to receive multiple copies of the direct path signal separately and compensate for the phase variations as in passive radar receivers.

\item [Backscatter propagation -] In backscatter case, the pathloss is inversely proportional to the fourth power of the utilized carrier frequency which severely limits the range of the system at higher frequencies. Moreover, in AmBC bi-static (AmBC-BS) case where both backscatter transceivers are powered by ambient signals (as shown in Figure~\ref{fig:ambc-bs}), the pathloss is inversely proportional to the product of the square distances between the transmitter (energy source) and the device, and the device and the receiver. If one of the distances is short, the other can be long as discussed in~\cite{Duan2020}. 

\item[Energy harvesting -] Even though the energy consumption of AmBC devices is much less than that of the active transmitters, it still needs some power to operate. The device could use battery, but better design is to use energy harvesting. Energy harvesting can also be utilized to make battery-free active devices~\cite{Lu2018}. 

\item[Co-existence with the legacy receiver -] Interference from AmBC devices to legacy receivers depend on the utilized waveforms. In case of OFDM based legacy system, AmBC causes interference only if its symbol duration is short compared to the OFDM symbol duration. Otherwise the AmBC modulated signal path appear as an additional multi-path component that can be tracked by the receiver~\cite{Ruttik2018}. AmBC device can also avoid interfering with the legacy receiver by shifting the scattered signal to the guard or adjacent band.
\end{description}


\subsubsection{Energy optimized Communication}

\subsubsection*{Zero-Energy and Wake Up Radios}

MTC devices significantly benefit when the battery-life matches useful lifetime (up to $\sim 40$ years) of the device. To accomplish this, RF front-ends may operate in two different modes – conventional low power active mode, and an always-on (even during sleep) very low power consumption power saving mode (PSM) running specific processes to capture, sense and classify the available signals, including wake up signals (WUS).

A true Zero-Energy (ZE) wake up radio (WUR) architecture consists of a specific ULP receiver path used to decode the information contained in the WUS frame and supplemented by an energy harvester front-end path for remote powering capabilities. Its adoption must consider supplementing the legacy air interface and its PHY layer (waveform and signaling) and system-level operational paradigms required by both the ULP receiver and EH. 

The power is carried by the optimized preamble of the frame that can be as simple as a single unmodulated RF carrier or a multi-tone power optimized waveform. Energy harvested from the preamble shall be sufficient to decode and operate the information contained in the WUS control and body fields.

A self-powered always-on feature can also be considered, for which EH from the incoming frame is not used. Thanks to adequate frame decoding, reachability of the MTC via paging with negligible delay is done on downlink signaling recognition. This can also make use of short duty cycling to further save power. 

Whatever true Zero-Energy or self-powered always-on, this WUS capture, initially introduced in 3GPP release 15 for NB-IoT and MTC devices and being further enhanced in release 16~\cite{3gpp36300}, necessitate a receiver embedding adaptive blocks that can adapt themselves to the current context. Requirements for sensing quite wide frequency bands to catch the current-in-use channels while maintaining correct decoding in the presence of strong signals is expected~\cite{maman2018}. Indeed, power consumption may benefit from event-driven blocks, coupled with digital supervising hardware designed with wake-up capabilities.

Discriminating signaling and information spread is supported from BSs side. It can also send individually- or group- addressed WUS frames, eventually with power optimized preambles within coverage for ZE MTCs.

\begin{figure}[htb]
    \centering
    \includegraphics[width=0.8\columnwidth]{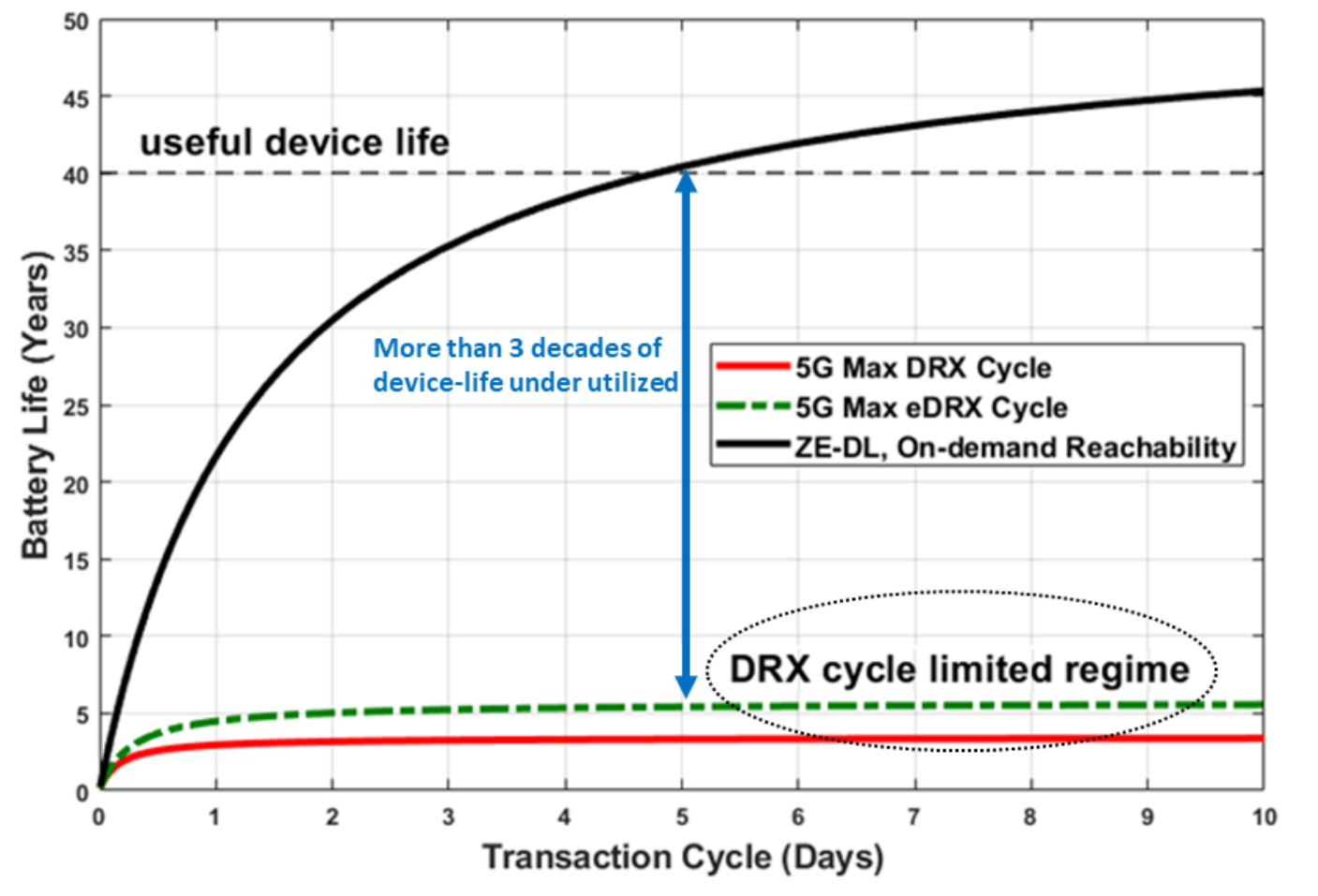}
    \caption{Battery life of a mobile IoT device using DRX-PSM and ZE-DL.}
    \label{fig:ZE_batteryLife}
\end{figure}

As an example, Figure~\ref{fig:ZE_batteryLife} shows MTD's battery life corresponding to legacy maximum DRX cycles of {$10.24$ s, $43.69$ min}, measurement cycle of $2.56$ s, and PSM on duty cycle of $25\%$. The figure further plots the MTD's battery life under the utilization of a ZE WUR. With the considered parameters, a ZE WUR operating over a new ZE air interface is capable of supporting $\sim 40$ years battery life – more than three decades of device life improvement over 5G.

Longer-term, the ZE air interface is expected to support the battery-less operation of IoT/ MTDs by eliminating the device's battery through the further enabling of limited payload data reception in the downlink direction as well as backscattering-based limited payload data transmission in the UL direction as discussed in the previous section.

\subsubsection*{Efficient hardware for on-device intelligence}

One of the key enablers for the next generation of MTDs, especially in the context of cMTC use cases, is the development of on-device intelligence blocks. The term intelligence here covers plethora of applications, usually based on artificial intelligence (AI) or machine-learning (ML) algorithms. These applications range from device adaptation regarding the operating context to save battery life, to smart Compressed Sensing schemes (CS), event-driven sensor interfaces or adaptive communication protocols to optimize data transfer~\cite{Chatterjee2019}. 

\begin{figure}[htb]
    \centering
    \includegraphics[width=0.8\columnwidth]{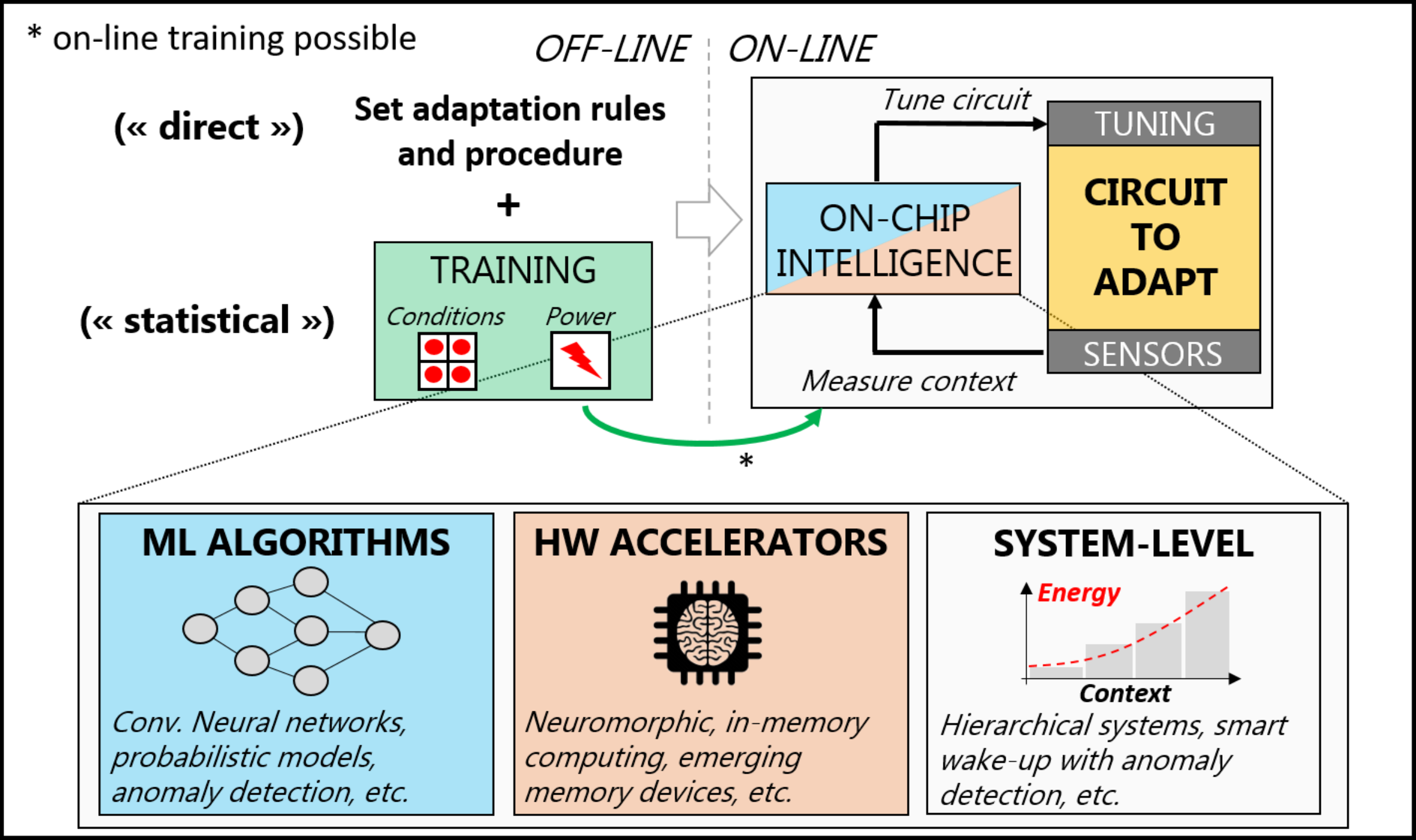}
    \caption{Illustration on possible strategies for on-chip device intelligence.}
    \label{fig:onChip}
\end{figure}

Embedding the required algorithmic functionalities on-chip while limiting the device's power consumption remains a challenge. Without PSM, dedicated processors are too power hungry for most embedded applications. Thus, an increased energy-efficiency means designing an integrated strategy, as illustrated in Figure~\ref{fig:onChip}. This includes (a) the ML algorithm with training (performed on-chip or off-chip), and (b) the hardware (HW) accelerators with its associated circuitry (sensors, tuning mechanisms, etc.).

For instance, small-scale tasks (e.g. smart-wake up) use low-power algorithms, such as probabilistic classifiers or decision trees. Large-scale tasks (e.g. on-device processing) use advanced algorithms, such as deep learning. In complement, hierarchical systems can be designed, adapting their performance and energy with the context (e.g. cascaded classifiers)~\cite{Goetschalckx2018}. 

Besides, HW accelerators are becoming neuromorphic, i.e. inspired by the energy-efficiency of the human brain. This means typically using analog processing blocks, event-driven circuits, and/or distributed memories~\cite{basu2018}. 


\subsection{Global Massively Scalable MTC}
\label{sub:scalable}

The success of MTC within the 6G ecosystem is tightly bound to the capability of offering globally available and massively scalable service for IoT solutions targeting a vast landscape of requirements. In this section, a number of fundamental technology enablers required to achieve this goal are highlighted. 

We start with some considerations on global coverage, focusing on the impact of frequency regulations and on the role of non-terrestrial networks, later to span the design implications along the protocol stack. The design of efficient non-orthogonal solutions capable of handling massive traffic over grant-free channels, possibly without channel state-information (CSI) is required at the physical layer, together with tailored channel code design. At the MAC layer, the peculiar traffic characteristics of MTC applications call for both novel random access schemes and advanced (persistent) scheduling approaches. Finally, we review the implications on higher layers.

\subsubsection{Global coverage}

\subsubsection*{Frequency Regulations}

The principal challenge obstructing implementation of global roaming for machines today is the discrepancy of regulations specifying the use of radio frequency spectrum in different countries. The frequency bands, the maximum transmit power, duty cycle and permitted media access mechanism(s), and even the allowed applications, may differ substantially from one region to another. This problem is especially vital for the unlicensed bands in-between 30~MHz and 2.4~GHz, but affects also cellular technologies, including 5G\footnote{\url{https://www.everythingrf.com/community/5g-frequency-bands}}. As a part of evolution towards 6G, moving to even higher frequency bands, which are today still mostly unregulated in many countries, provides the unique opportunity to harmonize spectrum worldwide and thus enable global roaming. However, a strong political effort will be needed to harmonize the spectrum in sub-GHz bands to support the machine applications inquiring long-range connectivity.

\subsubsection*{Non-terrestrial networks}

A key enabler towards true global connectivity for MTC is represented by non-terrestrial networks (NTN), leveraging the use of low-Earth orbit (LEO) satellites, drones, high-altitude platforms (HAPs) and unmanned aerial vehicles (UAVs) to dynamically offload traffic from the terrestrial system component and to reach otherwise unserved areas. First steps in this direction are already being taken within the 5G standardization, with a work- and study-item to support enhanced MBB (eMBB) and IoT services, respectively, starting from Release 17. In parallel, the market is witnessing the growth of commercial solutions providing global MTC connectivity via LEO constellations (e.g., very low power provider Astrocast\footnote{\url{https://www.astrocast.com/}}, high speed reliable Internet provider OneWeb\footnote{\url{https://www.oneweb.world}} and Amazon's project Kuiper), prompting NTN as a fundamental component in future of 6G~\cite{YA20_ruralConnectivity}. From this standpoint, new engineering challenges arise, e.g. to cope with high Doppler spread or to route information via inter-satellite or inter-drone links, calling for novel waveform and network designs.

\subsubsection{Physical Layer}

\subsubsection*{Non-orthogonal PHY solutions}

State-of-the-art solutions for efficient grant-free access allow multiple devices to share the same physical time and frequency resource, relying on distinguishable pilots. In the presence of mMTC, though, reliability may be jeopardized, and non-orthogonal multiple access (NOMA) is one of the key solutions to solve the resource collision issue. The success of this approach depends on both user detection and data decoding on the shared resources. 

With the periodic or sporadic traffic pattern of mMTC, only a small portion of devices are simultaneously active. Pilot detection can be stated as a compressed sensing problem, for which advanced algorithms like approximate message passing detection can be applied \cite{LLY_18_sparse}. 

On the other hand, advanced NOMA receivers call for a careful design of the multi-user detection (MUD) algorithm as well as iterative interference cancellation (IC) structure between MUD and channel decoders. Low complexity MUD and efficient IC structure to approach near maximum-likelihood performance while keeping the implementation cost acceptable is the key design philosophy. Examples like expectation propagation algorithms (EPA) with hybrid soft and hard IC structure have been proved to be efficient \cite{CBW_18_noma} in such cases. Joint user activity detection and decoding can also be further considered to optimize the performance.

While advanced NOMA has been well researched in configured grant-free schemes, in other grant-free approaches,\footnote{Under this general class we collect all schemes in which nodes transmit their messages without any type of pre-assignment or coordination.} the global power control, resource allocation and configuration cannot be accomplished efficiently, calling for advancements towards uncoordinated access policies. This poses the further challenge of multi-user interference (MUI), for which the one-dimensional randomness of the power domain yielded thanks to the near-far effect may not be enough. Instead, higher-dimensional randomness including also e.g. code- and spatial-domains should be introduced. 

In code domain schemes, prior knowledge of the statistical properties of data (e.g., constellation shape), codebook, and CRC should be fully utilized for advanced blind detection \cite{YHLD_18}. Although spatial domain NOMA is quite effective in improving the spectral efficiency, the use of conventional pilots to acquire channel information causes severe pilot contamination. Possible solutions include blind (pilot-free) data-driven methods~\cite{YLH_19_blind}, channel predictions using non-RF data~\cite{park2020extreme} and the enhancement of pilot design. The use of multi-pilots along with the application of strategies similar to modern random access to decode them is an example of the latter~\cite{yuan2020contentionbased}. Using the theory of modern random access, a number of users approaching the pilot length can be accommodated while resolving possible pilot collision via IC. 



\subsubsection*{CSI-free/limited schemes} 

CSI-based schemes allow compensating the channel impairments and consequently improve the communication performance. This holds when CSI acquisition costs is negligible, as in traditional broadband-orientated services. However, that may no longer be the case under strict constraints on latency, energy, and/or when serving a huge number of devices. In such cases, new and intelligent CSI-limited approaches are required. For instance, as network densification grows towards 6G era, the chances of operating under stronger line of sight (LoS) conditions increase, and beamforming schemes relying just on the channel statistics tend to reach near-optimum performance. Such statistical beamforming allows saving valuable resources in terms of energy and time since real-time CSI acquisition can be avoided.

\subsubsection*{Coding For Short Packets}

Error correcting codes employed in 5G NR include low-density parity-check (LDPC) codes and polar codes. The former has been tailored for the eMBB service class, whereas the latter are optimized for the transmission of control information. Both schemes require substantial adaptations to serve mMTC systems. In particular, bearing in mind the typical design choices of modern MAC protocols for mMTC, the following challenges can be identified: (i) the construction of close-to-optimal codes for intermediate block lengths (approx. from 100 bits to a few thousand bits); (ii) a strong error detection capability, possibly achieved without the overhead of an outer CRC code \cite{CDJLRSS_19}; (iii) the design of decoding algorithms capable to handle limited or no CSI at the receiver \cite{XCLOD_19}. The three identified challenges may lead to a modification of 5G coding schemes, by including profiles that are specifically addressing the mMTC setting.

\subsubsection{Medium Access}

\subsubsection*{Modern Random Access Schemes}

In the presence of a massive population of transmitters with intermittent and possibly unpredictable traffic pattern, classic scheduling approaches become rapidly inefficient, due to the explosion of the required overhead. Random access solutions appear as natural alternatives, providing the demanded flexibility. Unfortunately, classical ALOHA-based medium access schemes suffer from very limited efficiency thus impractical for the stringent demands of MTC in 6G. 

In the recent years, modern random access schemes have been proposed for answering such challenging quest~\cite{BCL_16_modernRA,Vem19:CompRA}. While the first schemes have been focusing on simple repetition of the transmitted messages and the use of interference cancellation at the receiver, more recently advanced multi-user code constructions and multi-user joint decoding have been investigated under the common umbrella of unsourced random access \cite{FJC19_SPARC}. A thorough investigation of the practical implications of using such very efficient schemes as a fundamental enabler for 6G MTC scenarios is still under study. Just to mention a few, how to enable user activity detection, user time-synchronization, and keep receiver complexity under control, are among the open problem in the research community. 

The presence of small cells and the rising interest in mega-constellation also opens the way for exploiting the presence of possibly many receivers to design efficient random access schemes \cite{Munari19_arXiv}.

\subsubsection*{Persistent Grant-Free Scheduling and Resource Allocation Design}

The heterogeneous traffic and QoS characteristics of mMTC requires tailored approaches to address the different applications. While modern random access solutions are promising for IoT devices with sporadic/unpredictable traffic, periodic data and strict time-dependent QoS characteristics such as jitter and latency of various MTC applications can be satisfied by persistent grant-free scheduling schemes. These solutions significantly reduce the control signaling burden and provide greater QoS-provisioning efficiency, and may be a strong candidate to achieve grant-free access for periodic data generating QoS-constrained massive MTC applications \cite{KGSC19}. 

An architecture in which sporadic, periodic and event-driven traffic are sliced into different bands, each supported by random-, persistent- or hybrid-access schemes can be envisioned for massive MTC towards 6G. The persistent scheduled access schemes can be jointly optimized exploiting PHY flexibility and multiple numerologies in order to satisfy diverse traffic and QoS requirements while achieving greater scalability \cite{SEP20}. Such cross layer optimization also needs to consider the coexistence among multiple numerologies, orthogonal/non-orthogonal waveforms etc., and can be efficiently addressed using artificial intelligence (AI) inspired solutions.


\subsubsection{Impact on higher layers}

\subsubsection*{Impact on higher layers of grant-free} 

\begin{figure}[htb]
    \centering
    \includegraphics[width=0.5\columnwidth]{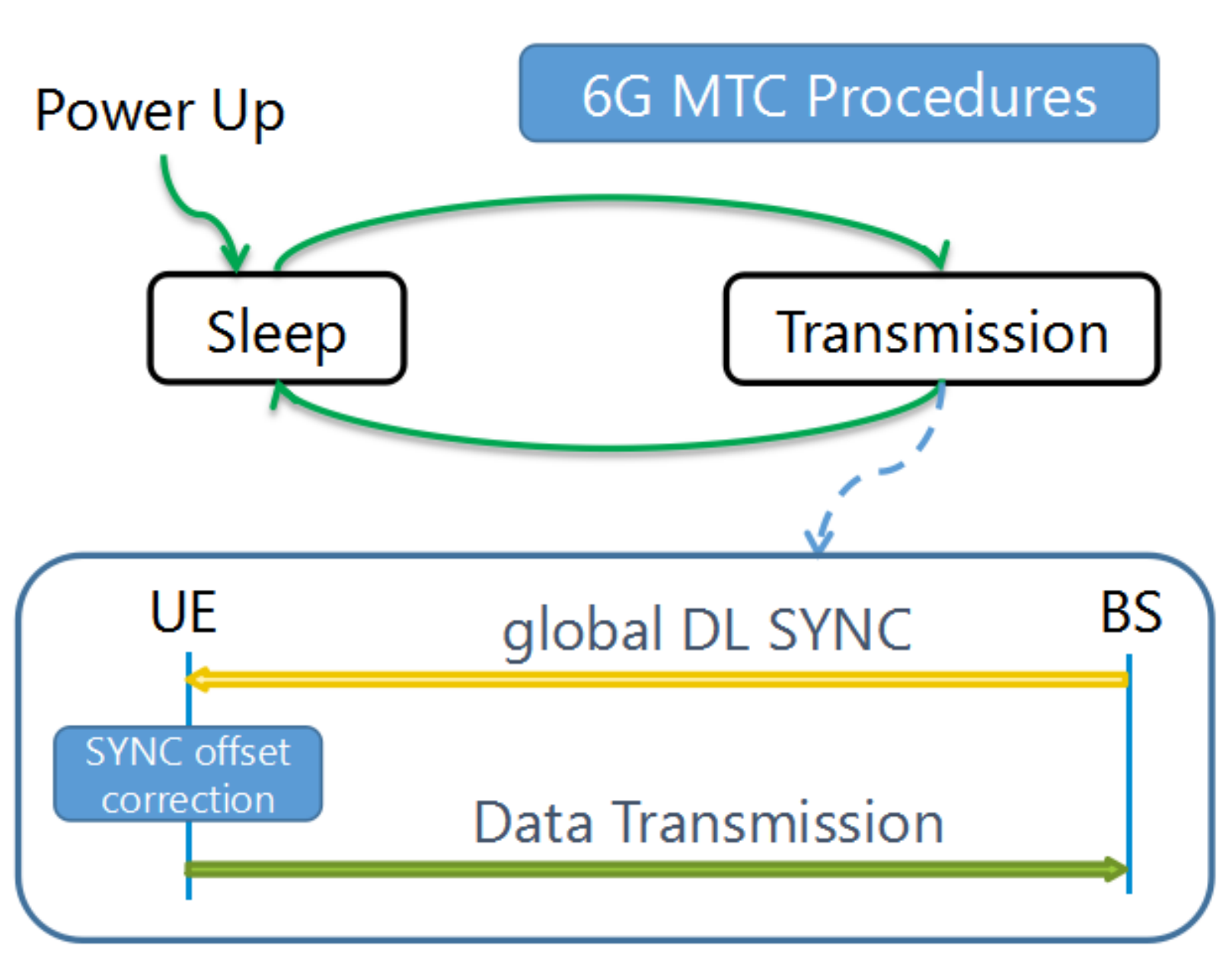}
    \caption{Simplification of higher layer operations.}
    \label{fig:higherLayer}
\end{figure}

When grant-free without any type of pre-assignment is employed, higher layers can be simplified without any connection state transition. How to effectively merge the multiple interactive procedures requires careful consideration. An example is shown in Figure~\ref{fig:higherLayer} where connection-free transmission can be achieved enabling instant MTC transmissions at any time without the delay of establishing a connection. 

However, the simplification also brings some challenges including synchronization (SYNC), privacy and security (cf. Section~\ref{sub:privacy}). To address the first issue SYNC offset correction should be used which can exploit the information of UE status (e.g., position, speed), base station (BS) position, SYNC signals of multiple BSs, etc. It also implies that the time offset and frequency offset correction should be further investigated when the pilots of different users are randomly superimposed.


\subsubsection*{Point-to-Multipoint capabilities in 6G RAN and core networks} 

Point-To-Multipoint (PTM) delivery is considered as a suitable transport mechanism for simultaneously delivering the same content to multiple devices within the covered area with a defined and stable QoS. While evolved Multimedia Broadcast Multicast Services (eMBMS) model can be considered as a successful trend in 4G LTE for applications such as video-on-demand, it has not been considered in 5G because of its inefficiency in terms of resource utilization and energy consumption. 

The support of PTM from the initial 6G design stages is therefore especially needed to address requirements of the forthcoming IoT deployments such as massive software updates. On the other hand, in current cellular IoT systems, the devices are still monitoring service announcements, even though firmware/software updates are rare. In that sense, novel on-demand paging methods would allow 6G IoT devices not to monitor service announcements but instead to be paged to receive multicast data, reducing energy consumption.

\subsection{Mission Critical MTC}
\label{sub:cMTC}

While 5G has already introduced mMTC for many IoT applications such as Smart City and Smart Home applications, we envisage that cMTC will be the primary focus of MTC in 6G. Those applications require dependable service quality characteristics in terms of latency and error rates, for example, in the context of life-critical alarming and control, practically equivalent to wired communications. In this sense, there is a close link with URLLC requirements with its envisaged target KPI values of a latency bound of $0.1$ ms combined with BLER of $10^{-9}$. 

\begin{figure}[htb]
    \centering
    \includegraphics[width=0.8\columnwidth]{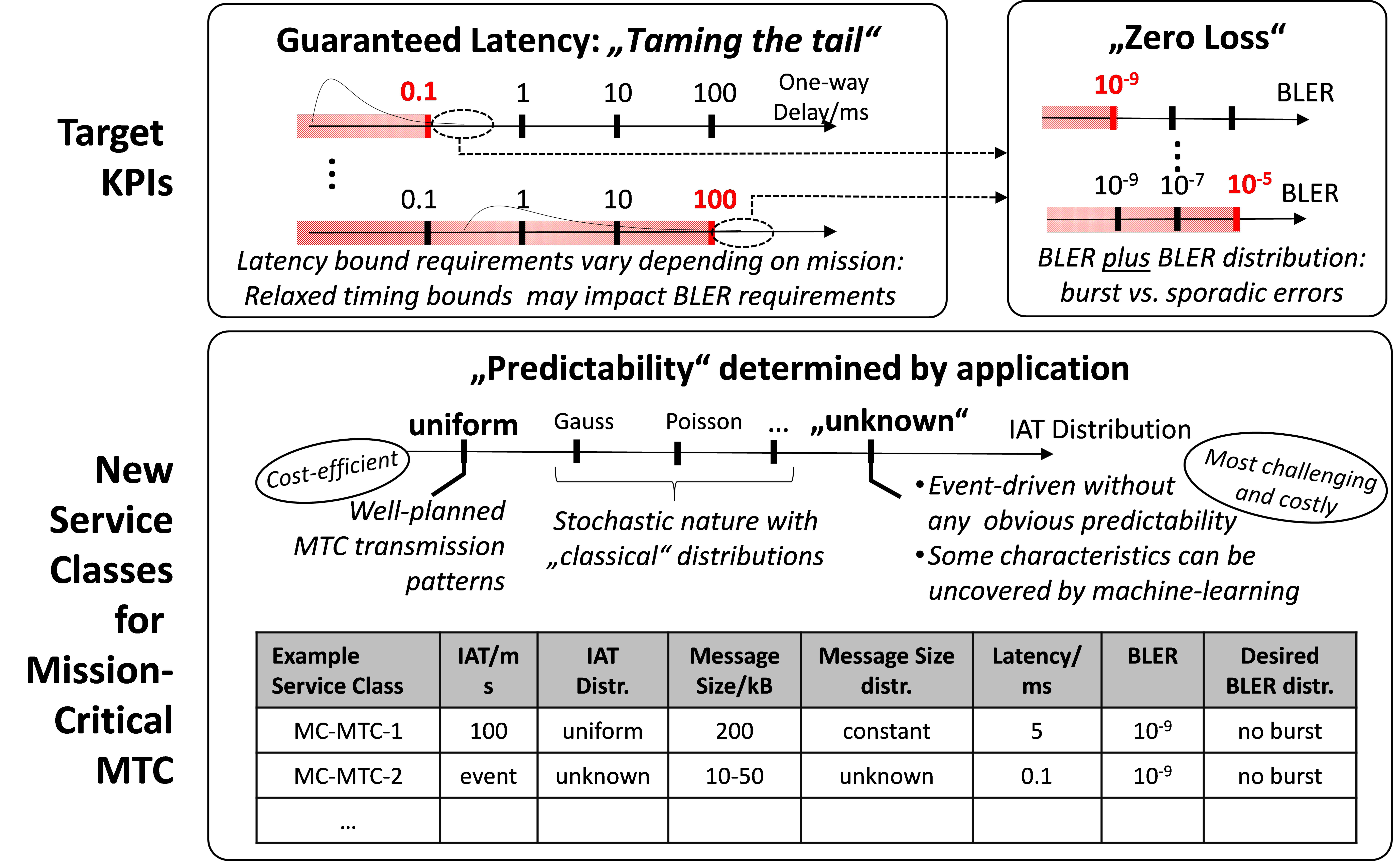}
    \caption{Mission-critical MTC challenge: \textit{taming the tail of latency and error distribution}.}
    \label{fig:cMTCchallenges}
\end{figure}

However, the extreme case of a very low absolute time boundary will only have practical relevance in a limited yet important number of use cases. In many cases, higher absolute end-to-end time bounds are acceptable, as long as the corresponding violation of the time-bound $-$ the ``taming of the tail'' $-$ as well as jitter are near zero~\cite{BDP18_urllc}. As depicted in Fig.~\ref{fig:cMTCchallenges}, some relaxation, both in terms of the absolute time-bound as well as the BLER and its distribution (burst vs. sporadic transmission), may be applicable to achieve resource-efficient and application-aware solutions~\cite{park2020extreme}. Mission-criticality also mandates a very high-security level, combined with resource efficiency required in an IoT environment (cf. Section~\ref{sub:privacy}).


The current 5G approach of tweaking the system design to meet URLLC requirements, for example, through shorter TTIs and data duplication via multi-connectivity, is neither scalable nor efficient in meeting the challenges of cMTC applications. For cMTC, future 6G systems should leverage application-domain information about the predictability of actual resource requirements and conditions: while ``classical'' network dimensioning had to consider the stochastic behavior of humans through corresponding Inter-Arrival Times (IAT) distributions of messages, the behavior of machine-type communications can be far more controlled and eventually even deterministic. Yet, especially event-driven, emergency-like MTC needs to be supported by 6G systems, with unknown knowledge of the IAT distribution and practically unpredictable behavior. The cost of achieving certain KPIs will be very different.

While regular transmissions can be efficiently scheduled within given time boundaries, scheduling of event-driven messages may require resource reservations and eventually lead to unused resources. Artificial intelligence can help to schedule algorithms to identify non-obvious regularities. Still, it might be a more effective way forward in 6G to allow cMTC applications to actively declare their transmission scheduling characteristics through newly introduced 6G cMTC service classes, each depending on ``classical'' parameters known from 5G, such as latency and BLER, but also on new parameters required for characterizing 6G requirements, such as predictability in terms of IAT distributions.

\begin{figure}[htb]
    \centering
    \includegraphics[width=0.8\columnwidth]{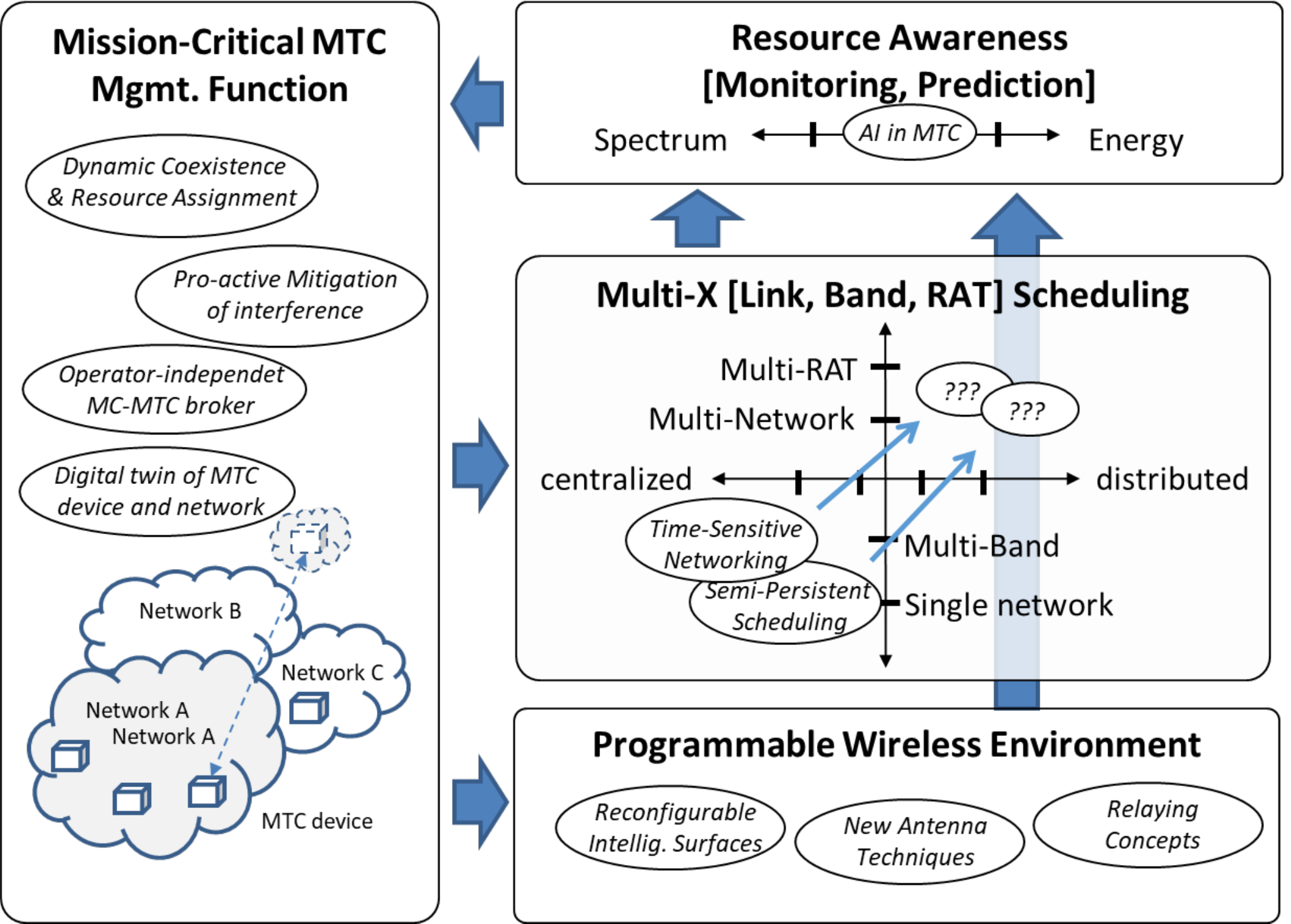}
    \caption{6G Mission-Critical MTC Solution Components}
    \label{fig:cMTCsolution}
\end{figure}

Based on those new cMTC-specific service classes, 6G systems need to allocate resources for mission-critical MTC appropriately within a multi-dimensional solution space comprising multi-RAT, multi-link, etc. In order to achieve solutions with acceptable cost, the absolute time bound needs to be chosen carefully and associated with a `price tag' in terms of spectrum usage and energy consumption. To enable such decisions in a heterogeneous, non-cellular-centric environment, a dedicated cMTC management function is needed. As illustrated in Fig.~\ref{fig:cMTCsolution}, this functionality considers resource awareness information, gathered from devices to control resource utilization of the networks (e.g., multi-RAT scheduling) and its environment (e.g., antennas and re-configurable intelligent surfaces (RIS)).


\noindent \textbf{The key building blocks for 6G cMTC are:}

\subsubsection*{Resource Awareness} 
The allocation of resources will require proactive monitoring of available resources and prediction of future resources for distributed user equipment as well as centralized network parts. Resource awareness should be supported by machine-learning and new network quality parameters delivered by the various networks (such as their current load level, which is an essential criterion for resource allocation, especially in distributed MTC networks)~\cite{SFW20_cooperative}.

\subsubsection*{The cMTC Radio Resource Management Function}
To facilitate mission-criticality in multi-network, multi-operator, programmable wireless environments, the MTC device - or a newly introduced broker functionality - needs to address the trade-off between the available and estimated resources across different link options (multi-RAT) and take the final decisions about the resource choices (e.g. scheduling, RIS). 

The cMTC broker function might be a new approach to off-load the resource-intensive decision process from the resource-constrained MTC device. The idea of a broker function has been adopted from cognitive networking concepts, in which a spectrum broker has been introduced to manage spectrum across different spectrum owners, see for example CBRS-Citizens Broadband Radio Service~\cite{KKY17_broker}. 

The broker function proposed to be introduced for cMTC does not only consider spectrum resources, but a broad range of radio resources, incl. scheduling options and the programmable wireless environment across different RATs. The broker function may also operate a digital twin of the critical MTC device in the field, and may act on behalf of it in order to not overload the resource-constraint MTC device. 

At the same time the digital twin allows for simulating and evaluating eventual decisions prior to their implementation. The idea of a digital twin corresponds well with architectural framework for machine learning in future networks outlined by ITU-T: in~\cite{ITU_3172} a sandbox environment is introduced ``in which machine learning models can be trained, tested and their effects on the network evaluated''.

\subsubsection*{Network Infrastructure and Wireless Environment Resources}
The key to achieve timing guarantees is the allocation of appropriate network resources, which can be allocated centrally or in a decentralized manner. Brute force approaches work but are highly inefficient, such as centrally assigned fixed resource reservations that might be actually used only very rarely. 

More flexible approaches such as semi-persistent scheduling and others are part of C-V2X and Time-Sensitive Networking, yet need to be evolved for 6G to work in multi-RAT and distributed environments. Finally, 6G is expected to offer not only network infrastructure components but also a programmable network environment to be controlled to serve cMTC devices.

\subsubsection*{Collision-friendly ultra-reliable transceiver design}
There are many resource allocation strategies to prevent collisions via AI prediction and semi-persistent scheduling. However, the collision cannot be fully prevented in some extreme cases where (1) the global information is missing, (2) the network topology varies very fast, and (3) the transmissions are massive and frequent, e.g. high-density  and high-mobility V2V in non-cellular domain. Therefore, collision-friendly transceiver design is required to ensure the reliability when collision happens. As the collision means the transmission is randomly non-orthogonal, truly grant-free NOMA methods jointly using spatial, code and power domain can be used to separate multiple collided users, and full duplex is required to ensure the receiving reliability when the cMTC device is transmitting~\cite{YMH_20_V2V}.

\subsection{Privacy and Security for MTC}
\label{sub:privacy}

 
Privacy and security aspects play a central role in any communication network. These include anomaly detection, (low-cost) authentication, data integrity and confidentiality and distributed trust, among others. Conventional solutions are not directly applicable to MTC networks owing to their fundamental differences, such as lack of human in the loop, massive deployment, diverse requirements spanning across the whole cost-complexity spectrum, and wide range of deployment scenarios. 

This section presents privacy and security related issues and enablers for MTC towards 6G. 
Privacy and data authorization are first presented, followed by a discussion on smart contract and blockchain technologies as distributed trust solutions. Finally, security issues and long-term secure data encryption and authentication techniques are detailed. 

\subsubsection{Privacy}

Though MTC has limited human involvement, there are some privacy threats that can be identified. In general, the threats are related to exposition of data related to individual's personal data in many of the identified use cases in this paper~\cite{ETSIref}. 
For instance, services related to autonomous mobility have person's location data (history data included), and possibly who they are with. In connected living use cases, the personal data is related to personal health and in ``Factories of the Future'' there may be employee related data. All of these data are under the European privacy legislation regulation (GDPR)\footnote{\url{https://gdpr-info.eu/}}. This necessitates privacy considerations when developing MTC based solutions.

Data flow and authorization management are well-engineered technology domains, where the modern solutions have been growing around OAuth: it is the building block for local trusted authorization (OAuth 2.0), identity provisioning (OpenID Connect) and user-centric, external authorization-as-a-service models (UMA - User Managed Access). 

Various OAuth profiles are also under active development for e.g. healthcare information exchange. There may be the need to develop a reference architecture for privacy management infrastructures where digitally signed consents enable personal data reuse between controllers under the individual's control, fulfilling the dynamic consent management related requirements of the GDPR when consent is used as the legal basis of personal data processing. One such solution might be the use of the MyData approach\footnote{\url{https://mydatafi.wordpress.com/}}.

\subsubsection{Smart Contracts and Blockchains as Enabling Concepts}

With a combination of features, blockchains are a broader way of looking at digital privacy and security. Privacy and security are central to the blockchain, including user identity security, transaction and communication infrastructure security, business security through transparency and audit as well as security from malicious insiders, compromised nodes or server failure~\cite{FS_blockchain}. 

Smart contract can be defined as ``a computerized transaction protocol that executes the terms of a contract''~\cite{szabo_smartContract} where contractual clauses are translated into code. This allows a system to embed the contract clauses to enforce the contract. 

Smart contracts, as part of blockchain technologies, offer a way to manage privacy and authentication by design. This provides a mechanism to establish decentralized trust by eliminating the need for a third party as a medium to guarantee the transaction while assuring privacy by technology~\cite{szabo_smartContract}. 

Smart contracts have become an integrated part of blockchain from the very beginning, allowing contracts to be stored as scripts and transactions to be executed inside the blockchain. Smart contracts are able to execute the contained code as they are triggered independently, allowing general purpose computation to occur. 

From a privacy perspective, a decentralized smart contract system like Hawk~\cite{KMS+16_hawk} does not store the financial transactions in public and thus ensures transactional privacy. The main advantage is that users can write a private smart contract in an intuitive manner without having to implement cryptographic protocols, which is accomplished by the Hawk compiler. 

On the other hand, Ethereum\footnote{\url{https://www.ethereum.org/}} allows users to write their own smart contracts and execute them in the blockchain. The approach taken by the Ethereum allows users to offer their services with smart contract, where the smart contract then communicates with other smart contracts in order to reach a contract. 

The key research area in smart contracts for MTC are mainly two-fold. Firstly, adopting smart contracts for resource constrained IoT devices introduces some inevitable technical challenges that mandates rethinking of existing blockchain/smart contact solutions. Secondly, MTC networks are conventionally uplink-oriented, with little to none peer-to-peer information exchange. Distributed trust requires two-way data exchange between devices, introducing new requirements and design challenges for 6G MTC networks.

\subsubsection{Security}

Security is only as strong as the weakest link and may change over time. Based on this principle it is natural to assume that security vulnerabilities may be identified in the life-cycle of any system. 

Even though the designers can anticipate most vulnerabilities by design, it is not possible to mitigate all zero-day vulnerabilities or the impact of zero-day exploits. To effectively tackle this security challenge the primary step is the identification of abnormalities in the system when they occur even if it is a zero-day vulnerability. 

Anomaly detection in software defined networks can be done using systems like SPHINX. However, the SPHINX implementation is not a comprehensive solution due to limitations in detecting transient attacks, packet integrity, malicious ingress or egress switches and so on~\cite{Sphinx}.

Another important aspect is the secure generation and exchange of cryptographic keys. A promising method are physical unclonable functions, which exploit physical fingerprints of devices that are inherently imposed by the manufacturing process to generate keys on-the-fly. Hence, cryptographic keys are rather generated on-the-fly than stored in a memory and thus are unknown to everyone, even the manufacturers.

\subsubsection{Long-term Secure Data Encryption and Authentication}

Conventional authentication, authorization and accounting processes are neither scalable nor cost-effective if directly applied to the large number of connected MTC devices. For example, the traditional subscriber identity module (SIM) based authentication solution is not suitable for massive deployment of low power and/or low cost MTDs. 

Lightweight and flexible solutions like group-based authentication schemes, anonymous service oriented authentication strategies to manage a large number of authentication requests, lightweight physical layer authentication, and the integration of authentication with access protocols represent promising solutions that are likely to be adopted in 6G. 

Another alternate is to use the unique properties of RF signatures to perform authentication. Machine learning techniques can be used to uniquely identify and authenticate MTDs by utilizing the inherent process variation on analog and RF properties of multiple wireless transmitters~\cite{CDM+19_RF_PUF}.

The long-term security, i.e. the protection of the confidentiality, authenticity and integrity of the transmitted and stored data, is another important aspect in MTC. In particular, the threat arising from attacks performed on quantum computers need to be considered~\cite{DAS_19_whatShould6G, 3GGP_post_quantum}. While the security of symmetric encryption schemes like the advanced encryption standard (AES) in the presence of quantum attacks can be recovered by adapting the key size, current asymmetric encryption, key-exchange and authentication schemes like Rivest-Shamir-Adleman (RSA) and Elliptic-Curve cryptography (ECC) based schemes can be compromised by quantum computing algorithms.

To secure the data in MTC in the age of quantum computing, lightweight and flexible quantum computer-resistant (or post-quantum) encryption and authentication schemes need to be considered. Currently, there is an ongoing standardization for quantum computer-resistant cryptosystems at the National Institute of Standards and Technology (NIST)~\cite{NISTpqc}.

Quantum key distribution (QKD) (also referred to as ``quantum cryptography'') is another direction towards ensuring the long-term security of data. The main difference between QKD and post-quantum cryptography is that the security of QKD is based on quantum effects, whereas post quantum cryptography aims at mitigating the threat given by quantum computers. Another important difference is that QKD requires an optical channel whereas post-quantum cryptosystems work with arbitrary channels.

Beside the use of computation-based security, such as AES, RSA, or ECC, the 6G MTC may also consider physical layer security. In this case, cryptographic keys are not needed which leads to information theoretic security, where the mutual information between the transmitted data and the data received by the eavesdropper $I(X; E)$ is minimized. This type of security can be achieved by the development of coding schemes based on Polar codes or Raptor codes.

\section{Conclusions}
\label{sec:outlook}

Following the trend of introducing a new cellular generation every decade, 2030s will witness the introduction of 6G. 5G services classes, namely URLLC, mMTC and eMBB, will be widely adopted and optimized with further enhanced requirements in 6G, while simultaneously introducing new use cases and service classes enabling digitalization of the society at large. MTC and IoT networks will form the main backbone of a 6G network providing wireless connectivity in all aspects of our everyday life.

This white paper provides an overarching view of MTC in the 6G era. The key drivers, potential use cases, evolving requirements and emerging service classes are first discussed. Future research directions considering different aspects of MTC considering both massive and critical MTC, ranging from the physical layer to the application layer and are then detailed in the rest of the paper. 

The key synopses of the discussion are synthesized in the six research questions presented in the Executive Summary at the beginning of this white paper.

\section*{Acknowledgements}
{\small This draft white paper has been written by an international expert group, led by the Finnish 6G Flagship program at the University of Oulu, Finland, within a series of twelve 6G white papers to be published in their final format in June 2020.}

\bibliographystyle{IEEEtran}
\bibliography{references.bib}

\begin{thebibliography}{10}
\providecommand{\url}[1]{#1}
\csname url@samestyle\endcsname
\providecommand{\newblock}{\relax}
\providecommand{\bibinfo}[2]{#2}
\providecommand{\BIBentrySTDinterwordspacing}{\spaceskip=0pt\relax}
\providecommand{\BIBentryALTinterwordstretchfactor}{4}
\providecommand{\BIBentryALTinterwordspacing}{\spaceskip=\fontdimen2\font plus
\BIBentryALTinterwordstretchfactor\fontdimen3\font minus
  \fontdimen4\font\relax}
\providecommand{\BIBforeignlanguage}[2]{{%
\expandafter\ifx\csname l@#1\endcsname\relax
\typeout{** WARNING: IEEEtran.bst: No hyphenation pattern has been}%
\typeout{** loaded for the language `#1'. Using the pattern for}%
\typeout{** the default language instead.}%
\else
\language=\csname l@#1\endcsname
\fi
#2}}
\providecommand{\BIBdecl}{\relax}
\BIBdecl

\bibitem{6gWP2019}
M.~Latva-aho and K.~Lepp\"{a}nen~(ed.), \emph{Key drivers and research
  challenges for {6G} ubiquitous wireless intelligence (white paper)}.\hskip
  1em plus 0.5em minus 0.4em\relax Oulu, Finland: 6G Flagship, Sep. 2019.

\bibitem{ITU_2410}
ITU, ``Minimum requirements related to technical performance for {IMT-2020}
  radio interface(s),'' Nov. 2017, report ITU-R M.2410-0.

\bibitem{GMB19:5Gevolution}
A.~{Ghosh}, A.~{Maeder}, M.~{Baker}, and D.~{Chandramouli}, ``{5G} evolution: A
  view on {5G} cellular technology beyond {3GPP} release 15,'' \emph{IEEE
  Access}, vol.~7, pp. 127\,639--127\,651, 2019.

\bibitem{KML18_6G}
M.~{Katz}, M.~{Matinmikko-Blue}, and M.~{Latva-Aho}, ``{6Genesis} flagship
  program: Building the bridges towards {6G}-enabled wireless smart society and
  ecosystem,'' in \emph{in Proc. IEEE 10th Latin-American Conference on
  Communications (LATINCOM)}, Guadalajara, Mexico, Nov. 2018.

\bibitem{david_6g_2018}
K.~David and H.~Berndt, ``{6G vision} and {requirements}: {Is} {there} {any}
  {need} for {beyond} {5G?}'' \emph{IEEE Vehicular Technology Magazine},
  vol.~13, no.~3, pp. 72--80, Sep. 2018.

\bibitem{SBC19_6G}
W.~{Saad}, M.~{Bennis}, and M.~{Chen}, ``A vision of {6G} wireless systems:
  Applications, trends, technologies, and open research problems,'' \emph{IEEE
  Network}, 2019.

\bibitem{ZXM_19_6G}
Z.~{Zhang} \emph{et~al.}, ``{6G} wireless networks: Vision, requirements,
  architecture, and key technologies,'' \emph{IEEE Vehicular Technology
  Magazine}, vol.~14, no.~3, pp. 28--41, Sep. 2019.

\bibitem{ZFW_19_6G}
B.~{Zong} \emph{et~al.}, ``{6G} technologies: Key drivers, core requirements,
  system architectures, and enabling technologies,'' \emph{IEEE Vehicular
  Technology Magazine}, vol.~14, no.~3, pp. 18--27, Sep. 2019.

\bibitem{SBG_19_6G}
E.~{Calvanese Strinati} \emph{et~al.}, ``{6G}: The next frontier: From
  holographic messaging to artificial intelligence using subterahertz and
  visible light communication,'' \emph{IEEE Vehicular Technology Magazine},
  vol.~14, no.~3, pp. 42--50, Sep. 2019.

\bibitem{viswanath6G_2020}
H.~{Viswanathan} and P.~E. {Mogensen}, ``Communications in the {6G} era,''
  \emph{IEEE Access}, vol.~8, pp. 57\,063--57\,074, Mar. 2020.

\bibitem{DAS_19_whatShould6G}
S.~Dang, O.~Amin, B.~Shihada, and M.-S. Alouini, ``What should {6G} be?''
  \emph{Nature Electronics}, vol.~3, no.~1, pp. 20--29, Jan. 2020.

\bibitem{MAL_20_6GMTC}
N.~H. Mahmood \emph{et~al.}, ``Six key enablers for machine type communication
  in {6G},'' in \emph{inProc. 2nd 6G Wireless Summit}, Levi, Finland, Mar.
  2020.

\bibitem{Nahavandi_I5dot0_2019}
S.~Nahavandi, ``Industry 5.0—a human-centric solution,''
  \emph{Sustainability}, vol.~11, no.~16, p. 4371, Aug. 2019.

\bibitem{Portilla.2019}
J.~{Portilla}, G.~{Mujica}, J.~{Lee}, and T.~{Riesgo}, ``The extreme edge at
  the bottom of the {Internet of Things}: A review,'' \emph{IEEE Sensors
  Journal}, vol.~19, no.~9, pp. 3179--3190, May 2019.

\bibitem{Elsts.2018}
\BIBentryALTinterwordspacing
A.~Elsts, E.~Mitskas, and G.~Oikonomou, ``Distributed ledger technology and the
  {Internet of Things}: A feasibility study,'' in \emph{Proceedings of the 1st
  Workshop on Blockchain-Enabled Networked Sensor Systems}, ser.
  BlockSys’18.\hskip 1em plus 0.5em minus 0.4em\relax New York, NY, USA:
  Association for Computing Machinery, 2018, p. 7–12. [Online]. Available:
  \url{https://doi.org/10.1145/3282278.3282280}
\BIBentrySTDinterwordspacing

\bibitem{5GACIA_connectedIdustries}
5G-ACIA, ``{5G} for connected industries and automation,'' Nov. 2018, 2nd. ed.

\bibitem{wentzloffULPsurvey}
\BIBentryALTinterwordspacing
D.~D. Wentzloff. (2020) Low power radio survey. [Online]. Available:
  \url{www.eecs.umich.edu/wics/low_power_radio_survey.html}
\BIBentrySTDinterwordspacing

\bibitem{Huynh2018}
N.~{Van Huynh} \emph{et~al.}, ``Ambient backscatter communications: A
  contemporary survey,'' \emph{IEEE Communications Surveys Tutorials}, vol.~20,
  no.~4, pp. 2889--2922, 2018.

\bibitem{Duan2020}
R.~{Duan} \emph{et~al.}, ``Ambient backscatter communications for future
  ultra-low-power machine type communications: Challenges, solutions,
  opportunities, and future research trends,'' \emph{IEEE Communications
  Magazine}, vol.~58, no.~2, pp. 42--47, 2020.

\bibitem{Lu2018}
X.~{Lu} \emph{et~al.}, ``Ambient backscatter assisted wireless powered
  communications,'' \emph{IEEE Wireless Communications}, vol.~25, no.~2, pp.
  170--177, 2018.

\bibitem{Ruttik2018}
K.~{Ruttik}, R.~{Duan}, R.~{Jäntti}, and Z.~{Han}, ``Does ambient backscatter
  communication need additional regulations?'' in \emph{2018 IEEE International
  Symposium on Dynamic Spectrum Access Networks (DySPAN)}, Seoul, South Korea,
  Oct. 2018, pp. 1--6.

\bibitem{3gpp36300}
{3GPP}, ``Evolved universal terrestrial radio access {(E-UTRA)} and evolved
  universal terrestrial radio access network {(E-UTRAN)}; overall description;
  stage 2,'' Jan. 2020, 36.300, v. 16.0.0.

\bibitem{maman2018}
M.~{Maman} \emph{et~al.}, ``Benefits of joint optimization of tunable wake-up
  radio architecture and protocols,'' in \emph{2018 25th IEEE International
  Conference on Electronics, Circuits and Systems (ICECS)}, Bordeaux, FRANCE,
  Dec. 2018, pp. 789--792.

\bibitem{Chatterjee2019}
B.~{Chatterjee}, N.~{Cao}, A.~{Raychowdhury}, and S.~{Sen}, ``Context-aware
  intelligence in resource-constrained {IoT} nodes: Opportunities and
  challenges,'' \emph{IEEE Design $\&$ Test}, vol.~36, no.~2, pp. 7--40, Apr.
  2019.

\bibitem{Goetschalckx2018}
K.~{Goetschalckx} \emph{et~al.}, ``Optimized hierarchical cascaded
  processing,'' \emph{IEEE Journal on Emerging and Selected Topics in Circuits
  and Systems}, vol.~8, no.~4, pp. 884--894, Dec. 2018.

\bibitem{basu2018}
A.~{Basu} \emph{et~al.}, ``Low-power, adaptive neuromorphic systems: Recent
  progress and future directions,'' \emph{IEEE Journal on Emerging and Selected
  Topics in Circuits and Systems}, vol.~8, no.~1, pp. 6--27, Mar. 2018.

\bibitem{YA20_ruralConnectivity}
E.~{Yaacoub} and M.~{Alouini}, ``A key {6G} challenge and opportunity —
  connecting the base of the pyramid: A survey on rural connectivity,''
  \emph{Proceedings of the IEEE}, vol. 108, no.~4, pp. 533--582, Apr. 2020.

\bibitem{LLY_18_sparse}
L.~{Liu} \emph{et~al.}, ``Sparse signal processing for grant-free massive
  connectivity: A future paradigm for random access protocols in the internet
  of things,'' \emph{IEEE Signal Processing Magazine}, vol.~35, no.~5, pp.
  88--99, Sep. 2018.

\bibitem{CBW_18_noma}
Y.~{Chen} \emph{et~al.}, ``Toward the standardization of non-orthogonal
  multiple access for next generation wireless networks,'' \emph{IEEE
  Communications Magazine}, vol.~56, no.~3, pp. 19--27, Mar. 2018.

\bibitem{YHLD_18}
Z.~Yuan, Y.~Hu, W.~Li, and J.~Dai, ``Blind multi-user detection for autonomous
  grant-free high-overloading multiple-access without reference signal,'' in
  \emph{In Proc. 87th Vehicular Technology Conference (VTC-Spring)}, Porto,
  Portugal, 2018.

\bibitem{YLH_19_blind}
Z.~{Yuan} \emph{et~al.}, ``Blind multi-user detection based on receive
  beamforming for autonomous grant-free high-overloading multiple access,'' in
  \emph{2019 IEEE 2nd {5G} World Forum (5GWF)}, Sep. 2019, pp. 520--523.

\bibitem{park2020extreme}
J.~Park \emph{et~al.}, ``Extreme {URLLC}: Vision, challenges, and key
  enablers,'' \emph{arXiv:2001.09683 [cs.IT]}, 2020.

\bibitem{yuan2020contentionbased}
Z.~Yuan \emph{et~al.}, ``Contention-based grant-free transmission with
  independent multi-pilot scheme,'' \emph{arXiv:2004.03225 [cs.IT]}, 2020.

\bibitem{CDJLRSS_19}
M.~C. Co{\.{s}}kun \emph{et~al.}, ``{Efficient Error-Correcting Codes in the
  Short Blocklength Regime},'' \emph{Elsevier Physical Communication}, vol.~34,
  no.~6, pp. 66--79, 2019.

\bibitem{XCLOD_19}
M.~Xhemrishi \emph{et~al.}, ``List decoding of short codes for communication
  over unknown fading channels,'' in \emph{In Proc. 53rd Annual Asilomar
  Conference on Signals, Systems, and Computers}, Pacific Grove, USA, Sep.
  2019.

\bibitem{BCL_16_modernRA}
M.~Berioli, G.~Cocco, G.~Liva, and A.~Munari, ``Modern random access
  protocols,'' \emph{Foundations and Trends® in Networking}, vol.~10, no.~4,
  pp. 317--446, 2016.

\bibitem{Vem19:CompRA}
A.~Vem, K.~Narayanan, J.-F. Chamberland, and J.~Cheng, ``{A user-independent
  successive interference cancellation based coding scheme for the unsourced
  random access Gaussian channel},'' \emph{IEEE Transactions on Communcations},
  vol.~67, pp. 8258--8272, 2019.

\bibitem{FJC19_SPARC}
A.~{Fengler}, P.~{Jung}, and G.~{Caire}, ``{SPARCs and AMP} for unsourced
  random access,'' in \emph{2019 IEEE International Symposium on Information
  Theory (ISIT)}, Jul. 2019, pp. 2843--2847.

\bibitem{Munari19_arXiv}
\BIBentryALTinterwordspacing
A.~Munari, F.~Clazzer, G.~Liva, and M.~Heindlmaier, ``{Multiple-Relay Slotted
  ALOHA: Performance Analysis and Bounds},'' 2019. [Online]. Available:
  \url{http://arxiv.org/abs/1903.03420}
\BIBentrySTDinterwordspacing

\bibitem{KGSC19}
G.~Karadag, R.~Gul, Y.~Sadi, and S.~Coleri, ``{QoS}-constrained semi-persistent
  scheduling of machine type communications in cellular networks,'' \emph{IEEE
  Transactions on Wireless Communications}, vol.~18, no.~5, pp. 2737--2750,
  2019.

\bibitem{SEP20}
Y.~Sadi, S.~Erkucuk, and E.~Panayirci, ``Flexible physical layer based resource
  allocation for machine type communications towards {6G},'' in \emph{In Proc.
  2nd 6G Wireless Summit 2020}, Levi, Finnland, 2020.

\bibitem{BDP18_urllc}
M.~{Bennis}, M.~{Debbah}, and H.~V. {Poor}, ``Ultra-reliable and low-latency
  wireless communication: Tail, risk, and scale,'' \emph{Proceedings of the
  IEEE}, vol. 106, no.~10, pp. 1834--1853, Oct. 2018.

\bibitem{SFW20_cooperative}
B.~Sliwa, R.~Falkenberg, and C.~Wietfeld, ``Towards cooperative data rate
  prediction for future mobile and vehicular {6G} networks,'' in \emph{In Proc.
  2nd 6G Wireless Summit}, Levi, Finland, Mar. 2020.

\bibitem{KKY17_broker}
T.~Kokkinen, H.~Kokkinen, and S.~Yrj\"{o}l\"{a}, ``Spectrum broker service for
  micro-operator and cbrs priority access licenses,'' in \emph{In Proc.
  Cognitive Radio Oriented Wireless Networks (CrownCom)}, Lisbon, Portugal,
  Sep. 2017.

\bibitem{ITU_3172}
I.-T. Y.3172, ``Architectural framework for machine learning in future networks
  including imt-2020,'' Geneva, Switzerland, Jun. 2019.

\bibitem{YMH_20_V2V}
Z.~Yuan, Y.~Ma, Y.~Hu, and W.~Li, ``High-efficiency full-duplex {V2V}
  communication,'' in \emph{In Proc. 2nd 6G Wireless Summit}, Levi, Finland,
  Mar. 2020.

\bibitem{ETSIref}
\BIBentryALTinterwordspacing
``{Analysis of Security Solutions for the oneM2M System},'' \emph{ETSI
  Technical Report DTR/oneM2M-000008}, 2014. [Online]. Available:
  \url{https://www.etsi.org/deliver/etsi_tr/118500_118599/118508/01.00.00_60/tr_118508v010000p.pdf}
\BIBentrySTDinterwordspacing

\bibitem{FS_blockchain}
F.~. Sullivan, ``Blockchains - redefining cybersecurity in a digital
  environment,'' Apr. 2016, research Report.

\bibitem{szabo_smartContract}
\BIBentryALTinterwordspacing
N.~Szabo. (1994) Smart contracts. [Online]. Available:
  \url{http://www.fon.hum.uva.nl/rob/Courses/InformationInSpeech/CDROM/Literature/LOTwinterschool2006/szabo.best.vwh.net/smart.contracts.html}
\BIBentrySTDinterwordspacing

\bibitem{KMS+16_hawk}
A.~{Kosba} \emph{et~al.}, ``Hawk: The blockchain model of cryptography and
  privacy-preserving smart contracts,'' in \emph{2016 IEEE Symposium on
  Security and Privacy (SP)}, California, USA, May 2016, pp. 839--858.

\bibitem{Sphinx}
M.~Dhawan, R.~Poddar, K.~Mahajan, and V.~Mann, ``{SPHINX}: Detecting security
  attacks in software-defined networks.'' in \emph{Ndss}, vol.~15, 2015, pp.
  8--11.

\bibitem{CDM+19_RF_PUF}
B.~{Chatterjee}, D.~{Das}, S.~{Maity}, and S.~{Sen}, ``Rf-puf: Enhancing iot
  security through authentication of wireless nodes using in-situ machine
  learning,'' \emph{IEEE Internet of Things Journal}, vol.~6, no.~1, pp.
  388--398, Feb. 2019.

\bibitem{3GGP_post_quantum}
\BIBentryALTinterwordspacing
Ericsson, ``{The Impact of Quantum Computers and Post Quantum Cryptography},''
  \emph{ETSI Technical Report DTR/oneM2M-000008}, 2016. [Online]. Available:
  \url{http://www.3gpp.org/ftp/tsg_sa/WG3_Security/TSGS3_85_Santa_Cruz/docs/S3-161847.zip}
\BIBentrySTDinterwordspacing

\bibitem{NISTpqc}
\BIBentryALTinterwordspacing
N.~I. of~Standards and T.~(NIST). Post-quantum cryptography standardization.
  [Online]. Available:
  \url{https://csrc.nist.gov/Projects/post-quantum-cryptography/Post-Quantum-Cryptography-Standardization}
\BIBentrySTDinterwordspacing

\end{thebibliography}

\end{document}